\documentclass[pra,twocolumn]{revtex4-1}
\usepackage[a4paper,centering,hmargin=1.75cm,vmargin=2cm]{geometry}

\usepackage{chemformula} 
\usepackage[T1]{fontenc} 
\usepackage{amsmath,amssymb}
\usepackage{bm}
\usepackage{xspace}

\newcommand{\etal}{\textit{et al.\xspace}}

\newcommand{\ansigma}[1]{\hat{c}_{#1\sigma}^{}}
\newcommand{\crsigma}[1]{\hat{c}_{#1\sigma}^{\dagger} }

\begin{document}

\title[]
{The princess and the pea: on the outsized role of inter-layer ligands in copper-pyrazine antiferromagnets}

\author{E. P. Kenny}
\affiliation{School of Mathematics and Physics, The University of Queensland, Brisbane, Queensland, Australia}
\email{elisekenny@gmail.com}
\author{A. C. Jacko}
\affiliation{School of Mathematics and Physics, The University of Queensland, Brisbane, Queensland, Australia}
\author{B. J. Powell}
\affiliation{School of Mathematics and Physics, The University of Queensland, Brisbane, Queensland, Australia}

\begin{abstract}
We investigate the cause of exchange anisotropy in a family of copper-based, quasi-two-dimensional materials with very similar geometries. This family differs mainly in the inter-layer separation, but have very different magnetic interactions even within the basal plane. We use density functional theory and Wannier functions to parameterize two complimentary tight-binding models and show that the superexchange between the \ch{Cu^2+} ions is dominated by a through-space interaction between hybrid Cu-pyrazine orbitals centered on the copper atoms. We find no correlations between the strength of this exchange interaction with homologous geometric features across the compounds, such as Cu and pyrazine bond lengths and orientations of nearby counter-ions. We find that the pyrazine tilt angles do not affect the Cu-pyrazine-Cu exchange because the lowest unoccupied molecular orbital on the pyrazine is at a very high energy (relative to the frontier orbitals, which are Cu-based). We conclude that the anisotropy of magnetic interactions within this family of materials is largely unpredictable before the crystal structure is firmly established -- it is due to non-homologous geometric features such as the inter-layer organic ligands previously thought to be benign.
\end{abstract}

\maketitle

\section{Introduction}
Low-dimensional magnetic crystals are of fundamental importance because they display fascinating quantum phenomena, such as quantum spin liquids \cite{ContempPhys}, topological order \cite{ContempPhys}, fractionalized excitations \cite{ContempPhys}, and exotic phase transitions \cite{KT}. Magnets based on organic and organometallic compounds display a wide range of interesting behaviors including quantum spin liquids \cite{RPP}, valence bond solids \cite{RPP}, dipole order \cite{Drichko,JackoPRM}, quantum dipole liquids \cite{Drichko,JackoPRM}, mechanomagnetism \cite{KennyACIE}, and proximal superconducting states \cite{RPP}. However, controlling these behaviors in new materials has proved enormously challenging, in part due to the difficulties in predicting and controlling the structures of molecular crystals. Coordination polymers offer a potential route around this problem because they offer the ability to control the local structure around the magnetic atoms via well understood coordination chemistry and longer range packing via the choice of linking molecules,\cite{NevilleBook} making crystal engineering a realistic possibility \cite{Robson}.

\begin{figure}
	\centering
	\includegraphics[width=\columnwidth]{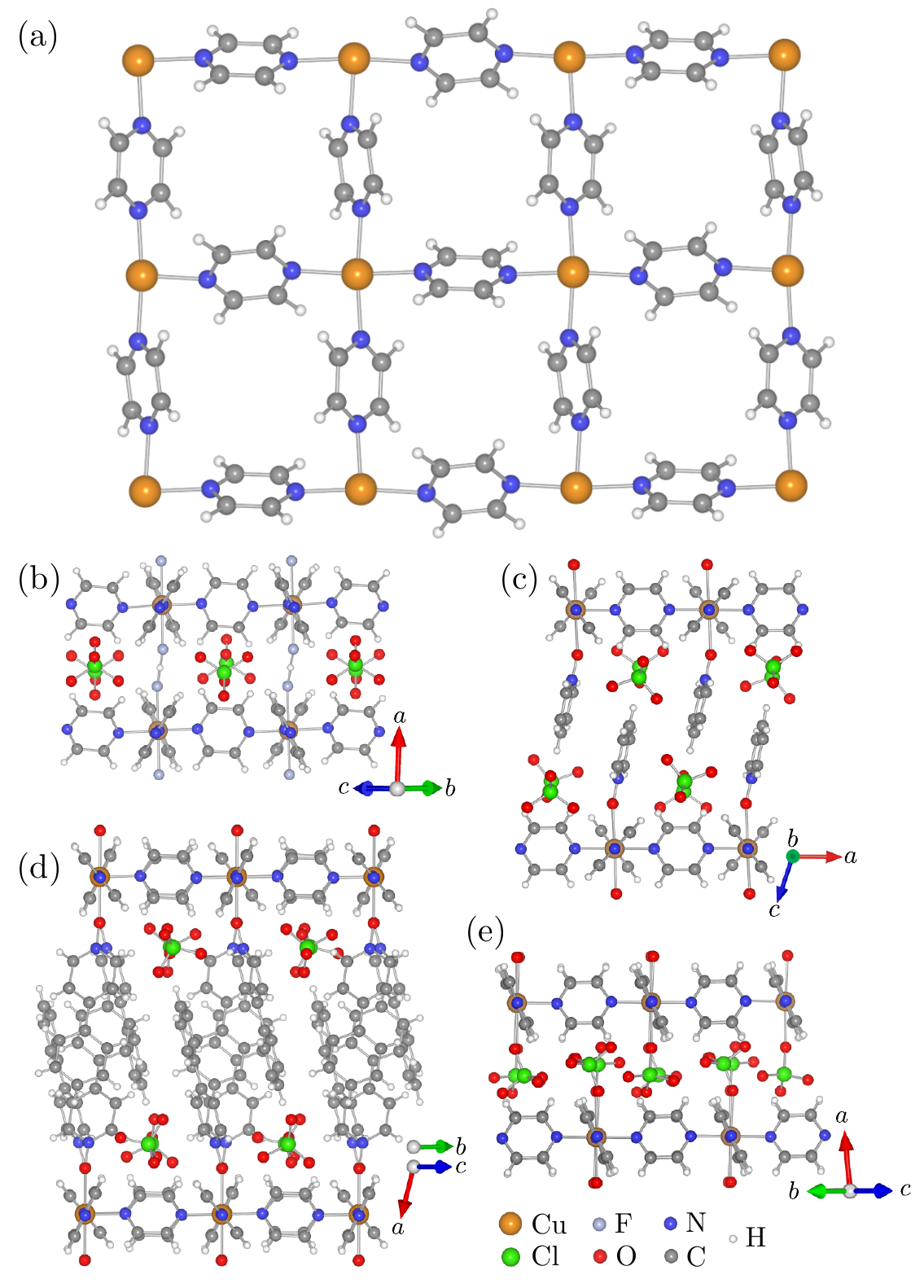}
	\caption{The four coordination polymers studied here\cite{Goddard2012}. (a) The square arrangement that is common to all four materials (here we show the structure of \textbf{1}). The other panels show the separation of the layers in (b) \textbf{1}, (c) \textbf{2}, (d) \textbf{3}, and (e) \textbf{4}. }\label{fig:goddard_structures}
\end{figure}

\begin{figure*}
	\centering
	\includegraphics[width=\textwidth]{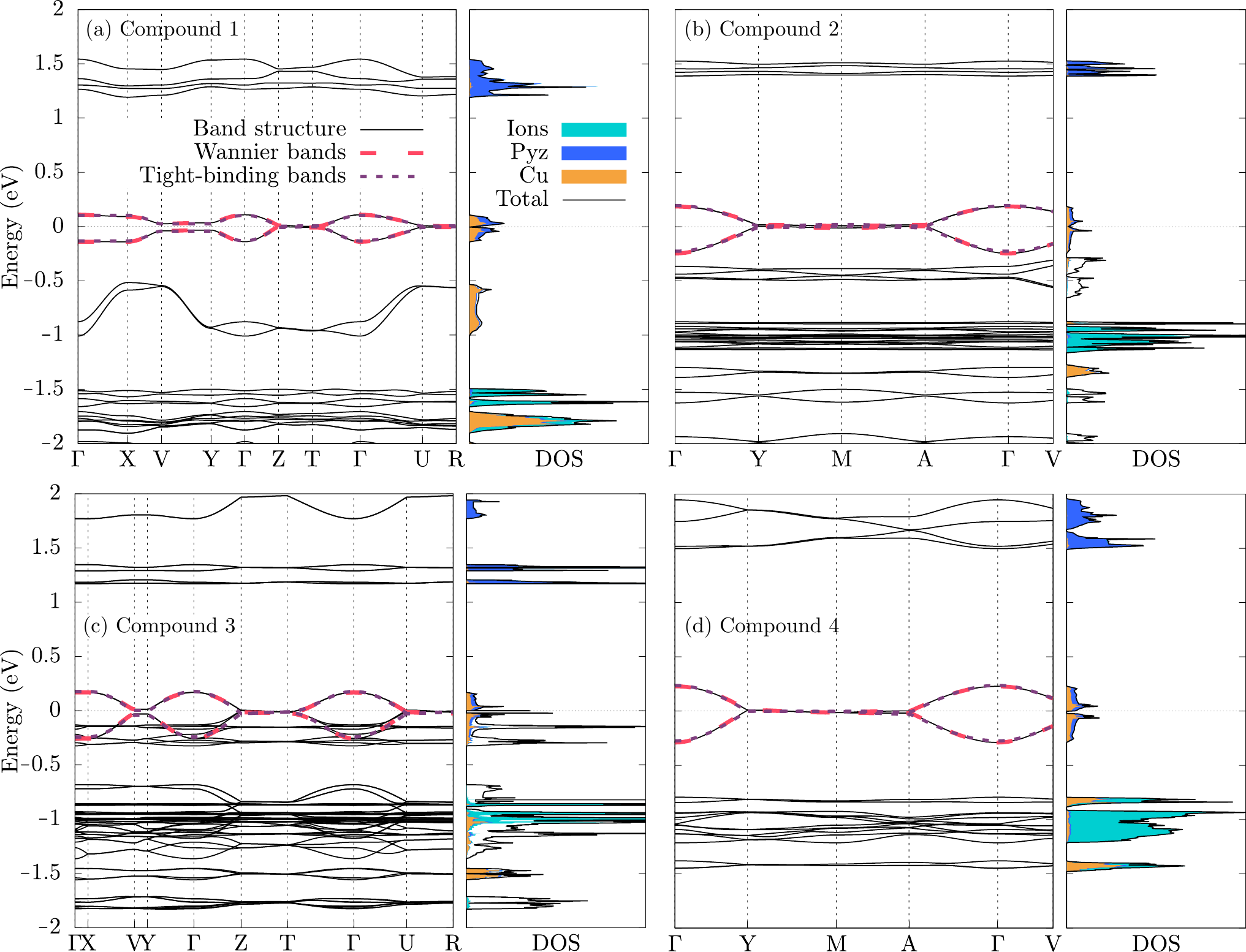}
	\caption{{\small Band structure, density of states (black line), and partial density of states (shading) for each compound. The two-Wannier tight-binding model reproduces the bands around the Fermi energy well. White areas in the partial density of states indicate contributions of all other atoms in the crystal (the inter-layer ligands).}}
	\label{fig:BS_DOS_goddard}
\end{figure*}

Copper-based coordination polymers have large potential for the realization of low-dimensional magnets \cite{Landee2013}. The exchange interactions in structurally homologous, chemically similar, materials can often vary widely \cite{Vela2013, Goddard2016}, offering an opportunity to design and investigate materials via finely tuning their geometries. Although an abundance of experimental information is available for these materials \cite{Tsyrulin2010, Tsyrulin2009, Darriet1979, Lancaster2007, Landee2013, Barbero2019, Barbero2016, DosSantos2016, Goddard2016, Goddard2012}, the factors that control the magnetic interactions are still debated. To be able to design materials with specific sets of magnetic interactions, these factors need to be understood in great detail.

In this work, we theoretically investigate a family of materials containing structurally homologous square-planar layers of \ch{Cu^2+} ions linked by pyrazine molecules (see Figure \ref{fig:goddard_structures}a). 
The layers are separated by different ligands in each material, coordinated to the \ch{Cu^2+} ion perpendicular to the planes. We examine four coordination polymers synthesized by Goddard \etal\ \cite{Goddard2016}. They are \footnote{We maintain the labeling used by Goddard \etal\ \cite{Goddard2016}.} 
(\textbf{1}) \ch{[Cu(HF2)(pyz)2]ClO4},
(\textbf{2}) \ch{[Cu(pyO)2(pyz)2](ClO4)2},
(\textbf{3}) \ch{[Cu(}4-phpy-O\ch{)2(pyz)2](ClO4)2},
and (\textbf{4})	\ch{Cu(pyz)2(ClO4)2},
where pyz = pyrazine, pyO = pyridine-N-oxide
, and 4-phpy-O = 4-phenylpyridine-N-oxide. 
All four materials, and many other \ch{Cu^2+} compounds with similar geometries, have been shown to behave as quasi-two-dimensional spin-1/2 Heisenberg antiferromagnets, with a weak superexchange via the pyrazine ligands \cite{Darriet1979, Lancaster2007, Tsyrulin2010, Landee2013}.

The details of pyrazine-mediated Cu-Cu superexchange, especially within the four materials studied here, have been under investigation and debate for some time \cite{Vela2013, DosSantos2016, Richardson1977, Hatfield1971, Mohri1999}. The anisotropy of the magnetic exchange interactions have been of particular interest. Broken-symmetry density functional theory calculations have shown that the magnitude of the superexchange between nearest-neighbor Cu atoms linked by pyz can vary by up to 200\% within a single compound, while the Cu-pyz-Cu bond lengths vary by a much smaller amount \cite{Vela2013, Goddard2016}. We aim to understand the origin of this anisotropy.

In the past, the superexchange anisotropy has been linked to the different dihedral angles of the pyrazine molecule -- being explained by a proposed superexchange pathway via the $\pi$-orbitals of the pyrazine \cite{Richardson1977}. Vela \etal\cite{Vela2013} have since shown (for \textbf{4}) that BS-DFT calculations give almost identical results when the pyrazine molecule is rotated. They proposed that the exchange anisotropy was caused by bond length alternation in the pyrazine ligand and the orientation of the nearby \ch{ClO4-} counter-ions. However, the pyrazine bond lengths in the other compounds studied here (especially \textbf{1}, which displays large differences in the in-plain nearest-neighbor interactions) differ by, at most, 0.001\,\AA, so this mechanism seems unlikely. Moreover, the experimental magnetic behavior of \textbf{4} has been shown to be only marginally affected by a shift in the orientations of the pyrazine rings and the \ch{ClO4-} counter-ions \cite{Barbero2019}.

Extended H{\"u}ckel calculations \cite{Mohri1999} and the investigation of experimental spin and charge densities \cite{DosSantos2016} have revealed an additional exchange pathway via the lone-pair $\sigma$-orbitals on the nitrogen atoms. Dos Santos \etal\cite{DosSantos2016} proposed, using the experimentally-determined bond ellipticity, that this $\sigma$-exchange mechanism may provide the biggest contribution to the superexchange pathway due to the absence of $\pi$ bonding interaction between \ch{Cu^2+} and pyrazine. In theory, both pathways ($\pi$ and $\sigma$) should be present in all compounds. 

\begin{figure*}
	\centering
	\includegraphics[width=\textwidth]{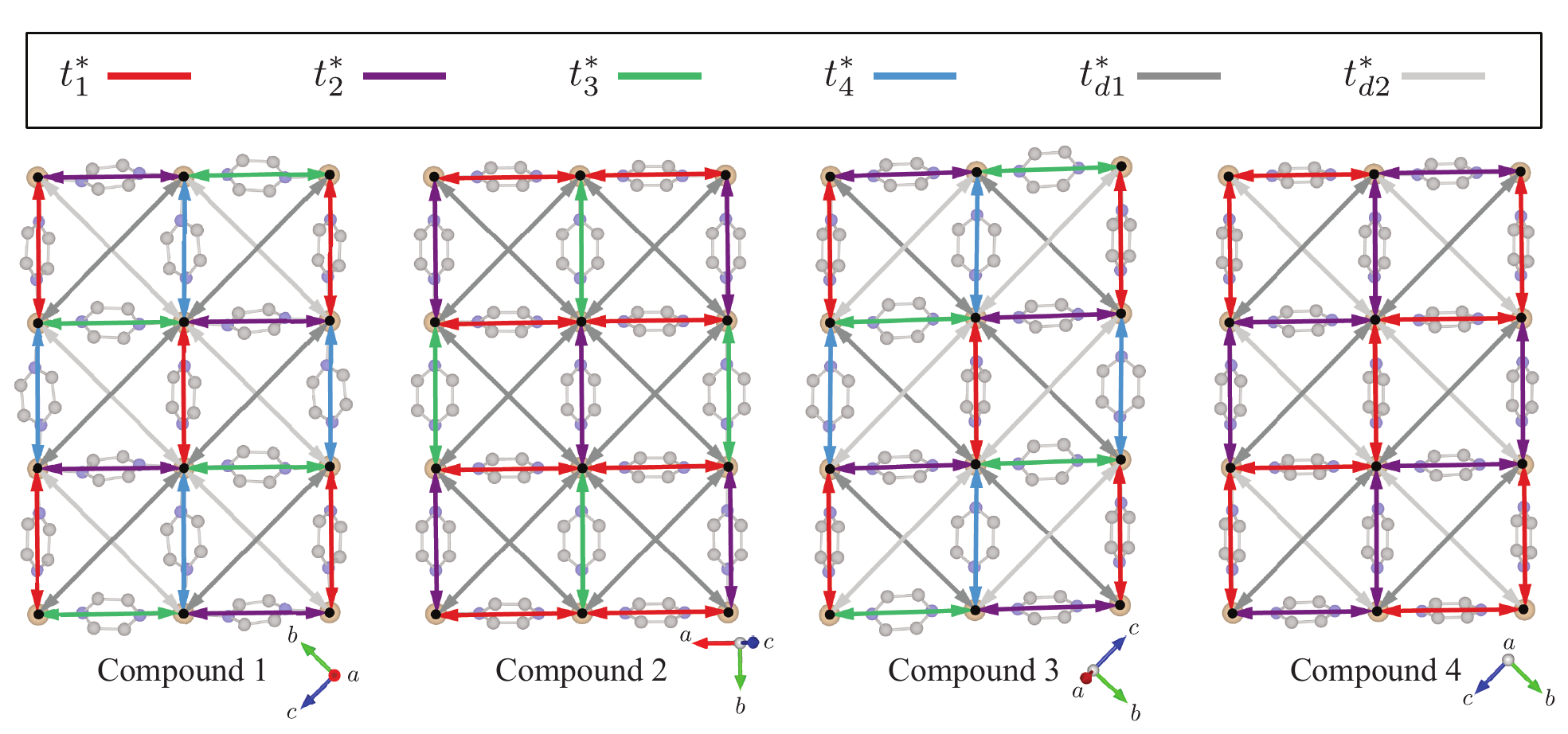}
	\caption{Effective Cu-pyz-Cu hopping integrals for each compound, shown in Tables \ref{tab:ts} and \ref{tab:upper_ts}.}
	\label{fig:ts}
\end{figure*}

Since the materials discussed here are structurally homologous, the only major difference being the inter-layer ligands,  we have the opportunity to identify structure-property relationships leading to changes in the magnetic superexchange interactions. We theoretically investigate the exchange mechanisms by parametrizing two tight-binding Hamiltonians for each compound; one with effective Cu-Cu hopping integrals and another with explicit Cu-pyz hopping. We also present density functional theory (DFT) calculations of the band structures and density of states. This allows us to provide a detailed explanation of the lack of $\pi$-mediated interactions and to narrow down the real cause of the superexchange anisotropy by searching for correlations between the hopping magnitudes and structural features. 

Kohn-Sham DFT is a single determinant theory. Therefore, it does not capture the strong electronic correlations, present in coordination polymers. This is evident, for example, as our DFT calculations predict metals whereas experimentally the materials studied here are Mott insulators \cite{Reimers-chapter}. Nevertheless DFT gives a material specific calculation of the one-electron terms in Hamiltonian, which allows us to calculate the hopping amplitudes, $t$, between localized (Wannier) orbitals. This provides insight into the microscopic origin of the superexchange interactions, $J$, as $J=4t^2/U$, where $U$ is the effect Coulomb repulsion on a Wannier orbital.

\section{Effective Cu-Cu Tight-Binding Model}

We construct an effective tight-binding model with two Wannier orbitals per unit cell (one per Cu atom),
\begin{equation}
\mathcal{H}_\mathrm{eff}=\sum_{\langle ij\rangle,\sigma}t^*_{ij}\left(\crsigma{i}\ansigma{j} + \crsigma{j}\ansigma{i}\right),
\end{equation}
where $\langle ij\rangle$ indicates a sum over nearest neighbour \ch{Cu^2+} sites $i$ and $j$ with the corresponding electron spins $\sigma$, and $t^*_{ij}$ are the effective electron hopping integrals between sites.

Since the \ch{Cu^2+} ions are separated by more than 6.5\,\AA\ in each compound, we expect that there is very little direct exchange between the sites. Thus, the tight-binding model can accurately describe the exchange interactions because they are most likely comprised almost entirely of superexchange as a result of virtual electron hopping.

Our DFT calculations were performed in an all-electron full-potential local-orbital basis using the FPLO package \cite{FPLO}. The electron densities were converged using a generalized gradient approximation (GGA) functional on a k-mesh of $7\times7\times7$. We perform calculations on the low-temperature crystal structures reported by Goddard \etal\cite{Goddard2016}. Figure \ref{fig:BS_DOS_goddard} shows the band structures and density of states for each compound. The partial density of states reveal, for all compounds, that the bands around the Fermi energy are comprised mostly of Cu and pyz orbitals, the upper bands of pyz orbitals, and the lower bands of counter-ion orbitals. We construct two Wannier functions per unit cell, each centered on a Cu atom. The bands closest to the Fermi energy are perfectly reproduced by our Wannier functions and the resulting tight-binding model.

\begin{figure*}
	\centering
	\includegraphics[width=0.9\textwidth]{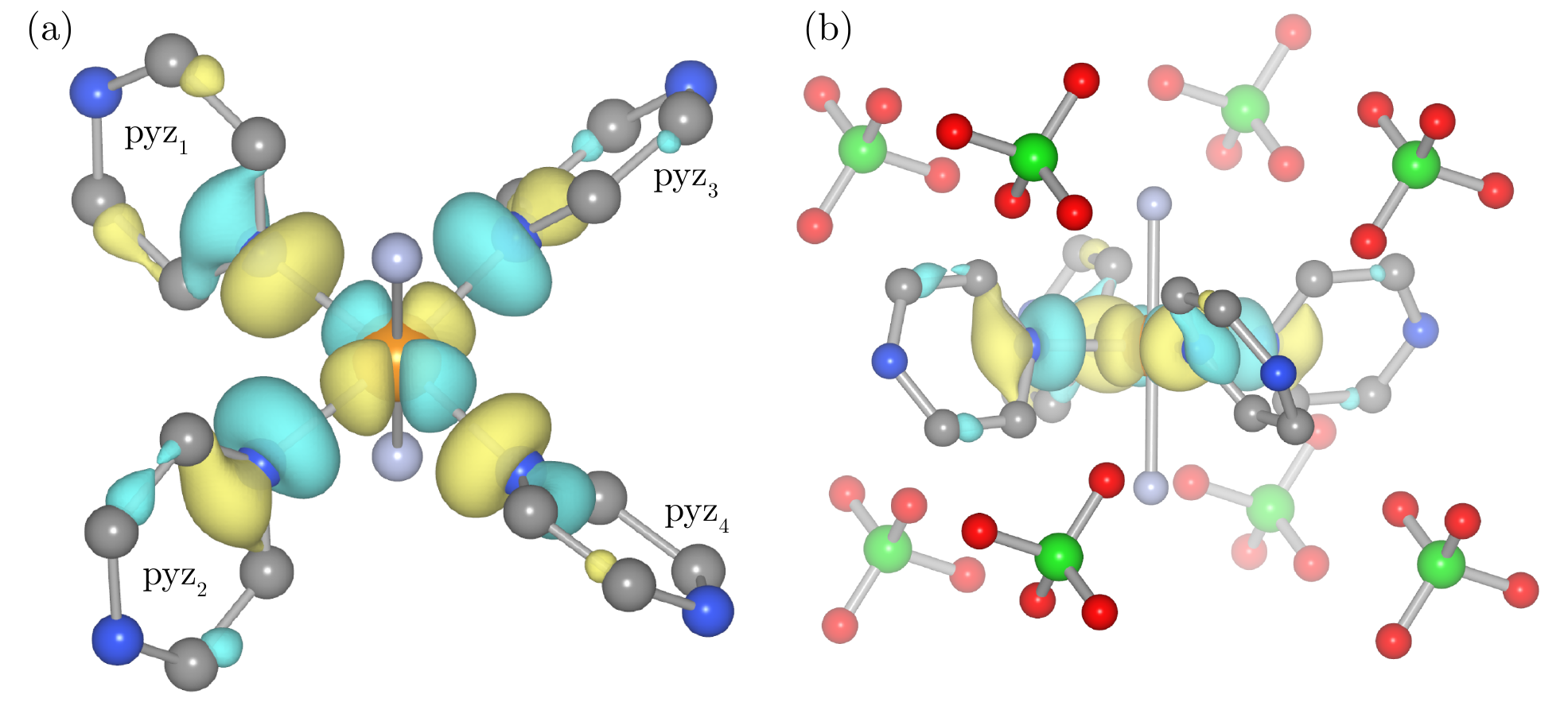}
	\caption{An isosurface plot of a Wannier function in \textbf{1} (a) and the positions and orientations of the counter-ions (b). It shows strong d$_{x^2-y^2}$ character on the Cu and strong sp$^2$ character on each N. The isosurface is at 0.02\,$e/$\AA$^3$. Wannier functions for the other compounds (shown in the Supporting Information) are very similar.}
	\label{fig:WF1}
\end{figure*}

Depending on the crystal symmetries, we find up to four distinct effective hopping interactions , $t^*_{ij}$, between Wanniers for each of the compounds. These are shown in Table \ref{tab:ts} and Figure \ref{fig:ts}, labeled in order from strongest to weakest. The largest intra-layer hopping disparity is found in \textbf{1}, where $t^*_1>2t^*_4$. In all compounds, the diagonal hoppings, $t^*_{d1}$ and $t^*_{d2}$, are small; we do not discuss them further in this work. Figure \ref{fig:WF1} shows a plot of a Wannier function in \textbf{1}, which is exemplary of those in the other compounds (included in the Supporting Information). It appears to be a superposition of the d$_{x^2-y^2}$ Cu orbitals and the sp$^2$ $\sigma$-orbitals on the N atoms. Visually, it is clear that the Wannier function is distributed more heavily on some pyrazine molecules compared to others. Those with the larger weights (pyz$_1$ and pyz$_2$ in the figure) correspond to the largest Cu-Cu electron hopping. The same is true for the other compounds, although the difference is not a stark as it is for \textbf{1}, since their hopping integrals are more uniform. Our Wannier functions are in good qualitative agreement with spin density and singly occupied molecular orbital plots found by others with BS-DFT for \textbf{4}\cite{Vela2013} and other similar compounds \cite{DosSantos2016}.

\begin{table}
	\begin{center}
		\caption{The tight-binding  parameters (meV) for the two-Wannier model calculated from  DFT. See Figure \ref{fig:ts}. A dash indicates that a parameter is not present  (due to high symmetry).}\label{tab:ts}
		\begin{tabular}{ccccccc} 
			\hline
			Compound & $t^*_1$ & $t^*_2$ & $t^*_3$ & $t^*_4$& $t^*_{d1}$& $t^*_{d2}$ \\ 
			\hline
			1 & 38.3 & 37.6 & 23.8 & 18.5 & -2.8 & -1.4 \\ 
			2 & 54.9 & 53.9 & 46.8 & - & -3.4 & - \\  
			3 & 63.2 & 44.8 & 43.4 & 42.4 & -3.5 & -1.9\\
			4 & 66.6 & 61.4 & - & - & -4.1 & -0.1 \\ 
			\hline
		\end{tabular}
	\end{center} 
\end{table}

\begin{figure}
	\centering
	\includegraphics[width=\columnwidth]{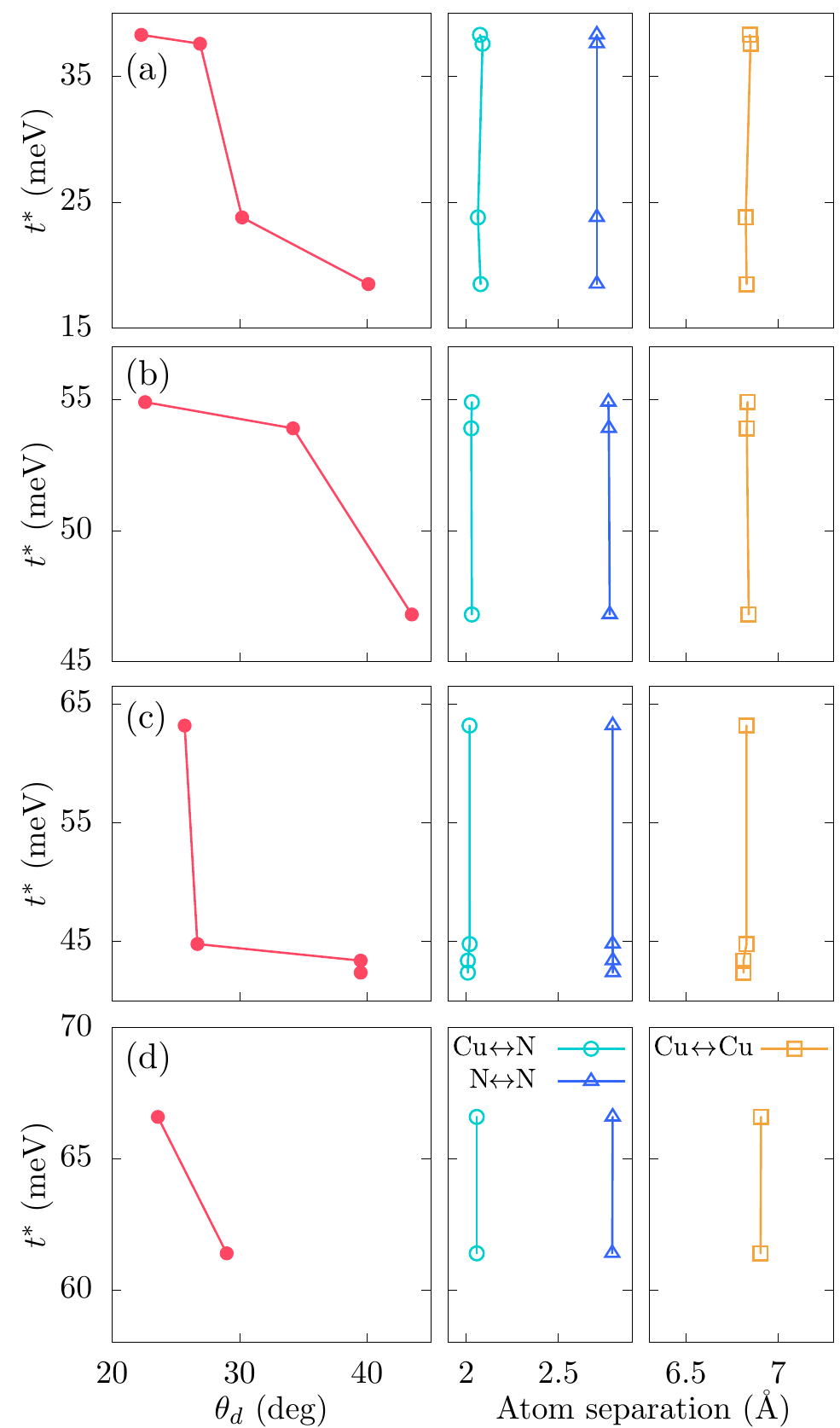}
	\caption{Dependence of Cu-pyz-Cu hoppings on various geometric parameters in (a) \textbf{1}, (b) \textbf{2}, (c) \textbf{3}, and (d) \textbf{4}. Although we find some correlation between the hopping magnitudes and the pyrazine twist angles, $\theta_d$, there appears to be no correlation with the various inter-atom distances; they remain constant for each Cu-pyz-Cu motif while the hopping parameters vary drastically. A table containing the data in this figure is included in the Supporting Information.} \label{fig:geom_v_t}
\end{figure}

Figure \ref{fig:geom_v_t} shows the dependence of Cu-pyz-Cu hoppings on various geometric parameters. Despite the large anisotropy in the hoppings within each compound, all atom separations are very similar. The most conspicuous geometric difference between each Cu-pyz-Cu exchange pathway is the pyrazine tilt angle, but as shown in Figure \ref{fig:geom_v_t} (and as observed in previous work \cite{DosSantos2016}), the maximum hopping does not coincide with the angle closest to 45$^\circ$, as one would expect if the effective hopping was largely influenced by the $\pi$-orbitals on the pyrazine bridge. 

\begin{figure}
	\centering
	\includegraphics[width=\columnwidth]{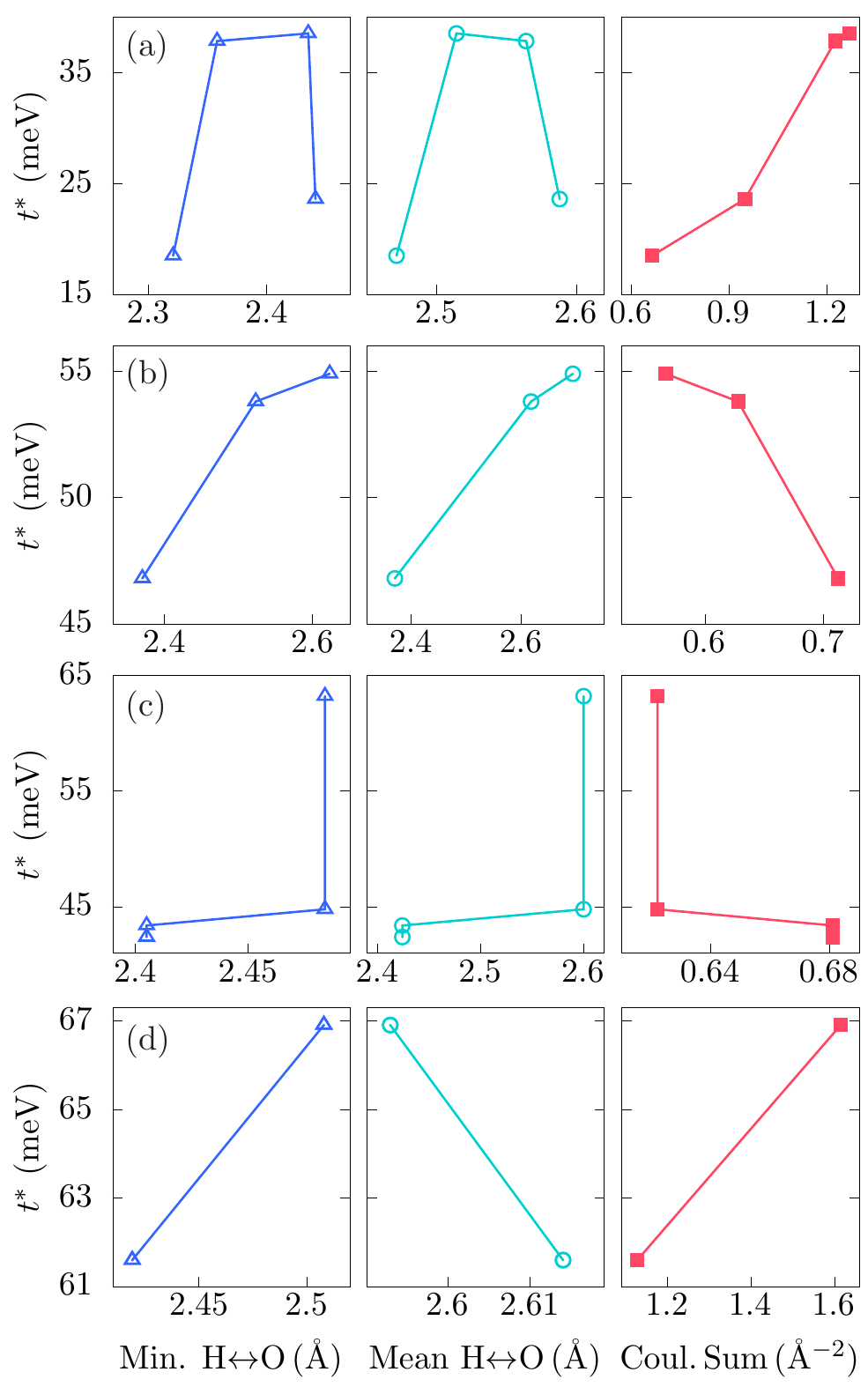}
	\caption{Dependence of Cu-pyz-Cu hopping on (from left to right) minimum H-O separation, the mean H-O separation, and the sum of nearest H-O Coulomb interactions according to Equation \ref{eq:coulomb_sum} for (a) \textbf{1}, (b) \textbf{2}, (c) \textbf{3}, and (d) \textbf{4}. Measurements are based on locations of H atoms on each pyrazine and O atoms on the nearest counter-ions. See Figure S1 in the Supporting Information for clarification on the distances used.}\label{fig:ion_stats}
\end{figure}

Vela \etal\ suggested that the orientations of nearby \ch{ClO-} counter-ions may have an effect on the Cu-pyz-Cu exchange, due to the possible interactions between the O atoms on the counter-ions and the H atoms on each pyrazine -- their relative orientation to each pyrazine is different, as shown in Figure \ref{fig:WF1}(b). Figure \ref{fig:ion_stats} shows the dependence of $t^*_{ij}$ on the minimum H-O separation, the average H-O separation, and the sum of nearest H-O Coulomb interactions,
\begin{equation}
	 \sum_\mathrm{n.n.} \frac{1}{|\bm{r}_\mathrm{H}-\bm{r}_\mathrm{O}|^2},\label{eq:coulomb_sum}
\end{equation}
where $\bm{r}_H$ and $\bm{r}_O$ are coordinates of nearby H and O atoms, respectively, and the sum is over all H and O atoms located between the center of each pyrazine and the Cl atom of the closest counter-ions (where the Cl atom is less than 6\,\AA\ away from the center of the pyrazine). See Figure S1 in the Supporting Information for clarification on the H and O pairs included in the sum. We find no consistent correlations between the effective hopping integrals and the proximity of nearby counter-ions across all compounds.  

\section{Considering Cu-pyz hopping explicitly}\label{Sec:Cu-pyz_hopping}

To investigate the electron hoppings between the copper atoms and pyrazine molecules, we define a new effective tight-binding model with six Wannier functions per unit cell (2 Cu-based and 4 pyz-based), where the hopping between any pair of Cu atoms is given by
\begin{equation}\label{eq:thru_ligand}
	\begin{split}
		\mathcal{H}_{i\sigma} =\ &t_{i\mathrm{a}} \left(\crsigma{1} \hat{c}_{l\sigma} +\hat{c}_{l\sigma}^\dagger\ansigma{1} \right) 
		+ t_{i\mathrm{b}}\left(\crsigma{2} \hat{c}_{l\sigma}+\hat{c}_{l\sigma}^\dagger\ansigma{2}\right)\\
		&+ t_i\left(\crsigma{2} \hat{c}_{1\sigma}+\hat{c}_{1\sigma}^\dagger\ansigma{2}\right)
		+ \frac{\Delta_i}{2}\left(\hat{n}_{l\sigma} - \hat{n}_{1\sigma} - \hat{n}_{2\sigma}\right),
	\end{split}
\end{equation}
where 1 and 2 label copper atoms and $l$ labels the lowest unoccupied molecular orbital (LUMO) of the pyrazine.

\begin{figure*}
	\centering
	\includegraphics[width=0.9\textwidth]{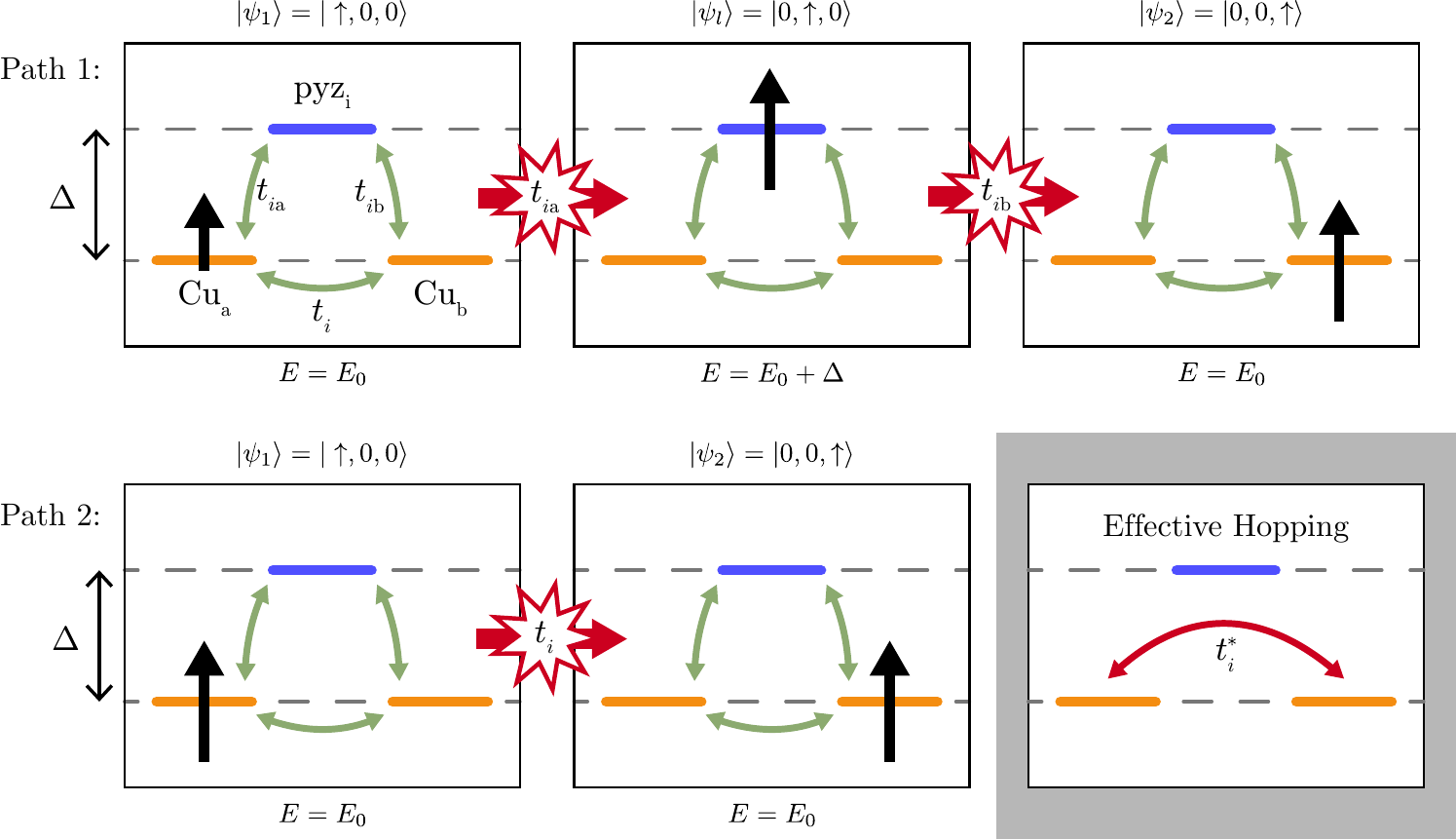}
	\caption{{\small The processes that give rise to the effective hopping of electrons between the \ch{Cu^2+} sites (shown in orange). The intermediate, unoccupied, pyrazine state (shown in blue) is higher in energy by $\Delta$. There are two paths that contribute to the overall effective hopping, $t^*_i$. Path 1 is via the unoccupied pyrazine state and Path 2 is a through-space interaction between the states localized on each \ch{Cu^2+} site. }} \label{fig:process}
\end{figure*}

Based on this six-Wannier model, we can derive the effective hopping $t^*_i$ as a function of the hoppings along the two paths (sketched in Figure \ref{fig:process}) using second order perturbation theory \cite{Reimers-chapter, GraafBroer}. It is
\begin{equation}\label{eq:eff_t}
	\begin{split}
		\tilde{t}^{*}_i &=\langle\psi_1|\mathcal{H}_{i\sigma}^\mathrm{eff}|\psi_2\rangle \\
		&= \langle\psi_1|\mathcal{H}_{i\sigma}|\psi_2\rangle - \frac{\langle\psi_1|\mathcal{H}_{i\sigma}|\psi_l\rangle\langle\psi_l|\mathcal{H}_{i\sigma}|\psi_2\rangle}{\Delta_i}\\
		&= t_i - \frac{t_{i\mathrm{a}}t_{i\mathrm{b}}}{\Delta_i},
	\end{split}
\end{equation}
where each of the variables are defined in Figure \ref{fig:process}.

\begin{figure*}
	\centering
	\includegraphics[width=\textwidth]{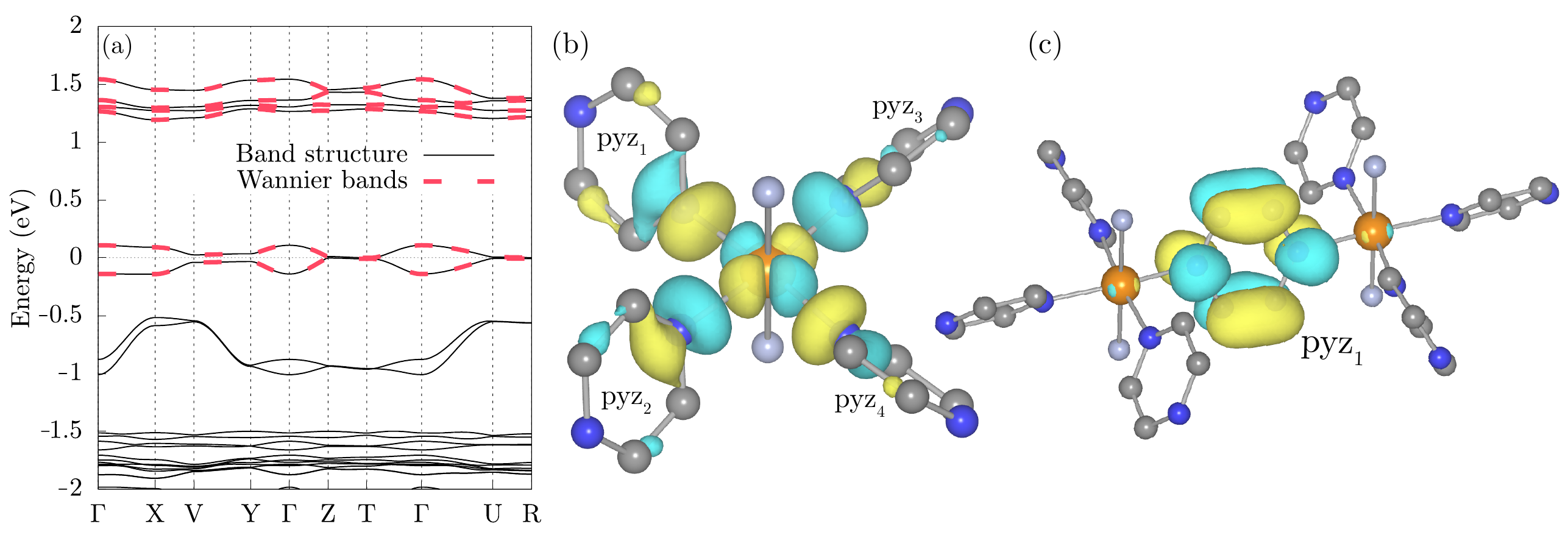}
	\caption{ The band structure (a) and isosurface plots (b), (c) for the six-Wannier model (2  Cu-based, 4 pyz-based) of \textbf{1} -- constructing additional Wanniers for the upper bands. The model, red, reproduces the DFT band structure well. (b) shows one of the Wannier functions centered on a Cu atom and (c) shows a Wannier function centered on a pyrazine molecule (in this case, pyz$_4$). The isosurfaces are at 0.02\,$e/$\AA$^3$.}
	\label{fig:WFpyz}
\end{figure*}

Using the same functional and k-mesh as in the previous section, we constructed additional Wannier functions localized on each pyrazine, corresponding to the upper bands in Figure \ref{fig:BS_DOS_goddard}. Figure \ref{fig:WFpyz} shows the resulting band structure and exemplar isosurface plots for the Wannier functions in \textbf{1} (band structures and Wannier function plots for the other compounds are supplied in the Supporting Information). In all compounds, the Wannier functions localized on each pyrazine molecule resemble the LUMO of an isolated pyrazine \cite{Xin2002}.

\begin{figure}
	\centering
	\includegraphics[width=0.8\columnwidth]{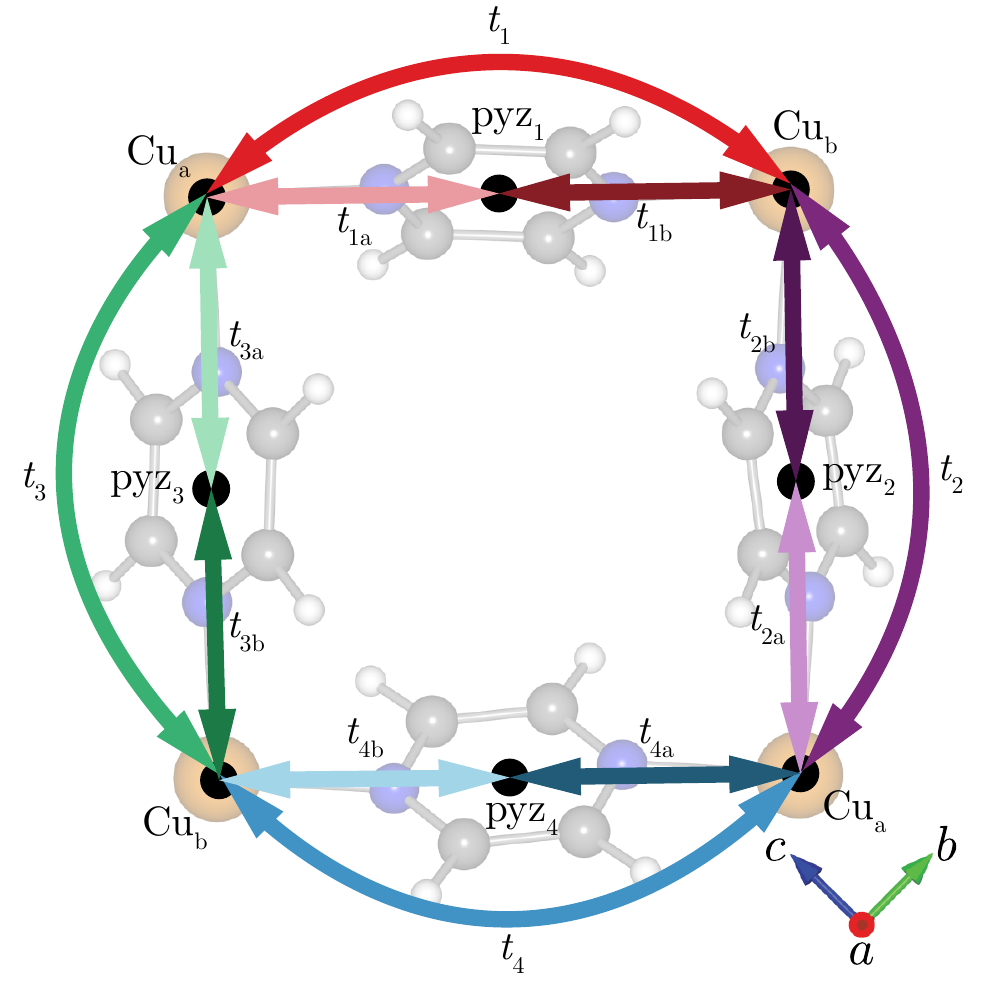}
	\caption{Hopping parameters included in the six-Wannier model, Equation \ref{eq:thru_ligand}. This is the in-plane structure of \textbf{1}; the extension to the other compounds is self-explanatory.}\label{fig:Cu_pyz_ts}
\end{figure}

Table \ref{tab:upper_ts}  shows our results for the new hopping parameters (see also Figure \ref{fig:Cu_pyz_ts}), the energy difference (between the Wannier functions on Cu and pyz), $\Delta_i$, and the resultant effective hoppings, $\tilde{t}^*_i$, according to Equation \ref{eq:eff_t}. The Cu-pyz hopping parameters ($t_{ia}$ and $t_{ib}$) are, in most cases, larger than the through-space hopping parameters ($t_i$). The effective hoppings $\tilde{t}^*_i$ agree well with the effective hopping $t^*_i$ calculated from the two-Wannier function construction (Table \ref{tab:ts}).

\bgroup
\def\arraystretch{1.2}
\begin{table*}
	\begin{center}
		\caption{The tight-binding parameters for the six-Wannier model  (2  Cu-based, 4 pyz-based) defined in Equation \ref{eq:thru_ligand}. The last column contains the effective hoppings from the model with two Cu-based Wannier functions (Table \ref{tab:ts} and Figure \ref{fig:geom_v_t}). See Figure \ref{fig:Cu_pyz_ts} for the definition of the labeling. }\label{tab:upper_ts}
		\begin{tabular*}{\textwidth}{c @{\extracolsep{\fill}} cccccccc}
			\hline
			Compound & $i$ & $t_i$ (meV) & $t_{i\mathrm{a}}$ (meV) & $t_{i\mathrm{b}}$ (meV) & $\Delta_i$ (eV) & $t_{i\mathrm{a}}t_{i\mathrm{b}}/\Delta_i$ (meV) & $\tilde{t}^*_i$ (meV) & $t^*_i$ (meV) \\ 
			\hline 
			1   & 1 & 38.4 & -8.3 & 8.3 & 1.39 & -0.1 & 38.5 & 38.3\\ 
			& 2 & 35.3 & -59.4 & 59.4 & 1.40 & -2.5 & 37.8 & 37.6\\ 
			& 3 & 23.1 & -24.5 & 24.5 & 1.31 & -0.5 & 23.6& 23.8\\ 
			& 4 & 17.3 & 30.3 & -30.3 & 1.25 & -0.7 & 18.0& 18.5\\
			\hline
			2   & 1 & 54.8 & -9.0 & 9.0 & 1.46 & -0.05 & 54.9 & 54.9 \\ 
			& 2 & 53.8 & $<10^{-7}$ & $<10^{-7}$ & 1.48 & 0.0 & 53.8 & 53.9 \\ 
			& 3 & 46.8 & $<10^{-7}$ & $<10^{-7}$  & 1.41 & 0.0 & 46.8 & 46.8 \\ 
			\hline
			3   & 1 &  61.7 & 32.9  & -32.9  & 1.33 & -0.8 & 62.5 & 63.2 \\ 
			& 2 &  32.1 & -131.2  & 131.2  & 1.30  & -13.2 &  45.3 & 44.8 \\ 
			& 3 & 43.1  & -13.7  & 13.7  & 1.21  & -0.2 &  43.3 & 43.4 \\ 
			& 4 &  40.6 &  39.1 & -39.1  & 1.21  & -1.3 & 41.9  & 42.4 \\ 
			\hline
			4  & 1 & 66.7 & -17.1 & 17.1 & 1.74 & -0.2 & 66.9 & 66.6\\ 
			& 2 & 60.7 & 39.2 & -39.2  & 1.64 & -0.9 & 61.6 & 61.4\\
			\hline
		\end{tabular*} 
	\end{center} 
\end{table*}
\egroup

For every effective hopping, the sign differences of the Cu-pyz hoppings lead to an overall increase in magnitude, since the second term in Equation \ref{eq:eff_t} is always negative. However, the separation of energy, $\Delta_i$, between the metal and ligand states is so large that the effect of the $\pi$-interactions in Path 1 (see Figure \ref{fig:process}), quantified by $t_{i\mathrm{a}}t_{i\mathrm{b}}/\Delta_i$, is relatively small. This explains why the Cu-pyz-Cu hopping is  weakly correlated with the orientation of the pyrazine molecule. $t_{ia}$ and $t_{ib}$, have a small effect due to the large energy gaps $\Delta_i$ and the through-space hoppings, $t_i$, dominate the interaction. The only exception to this is  $\tilde{t}^*_2$ in Compound 3, which is increased by 41\% on accounting for the $\pi$-interactions. In this case the energy separation remains large, but $t_{1a}$ and $t_{1b}$ are large enough to allow significant exchange through the pyz anyway. This means that the $\pi$-interactions in the Cu-pyz-Cu motif cannot be neglected in general.

The Wannier functions centered on the \ch{Cu^2+} in both of our models are extremely similar (comparing Figures \ref{fig:WF1}(a) and \ref{fig:WFpyz}(b)). This confirms that the weighting on each pyrazine is largely independent of the pyrazine LUMO orbitals and suggests that there is another cause of the disparity between the intra-layer hoppings. 

\section{Conclusion}

With the use of DFT and Wannier functions, we have parameterized two complimentary tight-binding models for four similar coordination polymers, all exhibiting Cu-Cu magnetic superexchange mediated by pyrazine ligands. We have shown that, in most cases, the effective Cu-pyz-Cu hoppings are dominated by a through-space interaction between hybrid Cu-pyz orbitals centered on the copper atoms. 

According to our findings, the anisotropy of the Cu-Cu magnetic superexchange in these compounds cannot be definitely linked to any variations in homologous geometric features across all compounds including the pyrazines, the counter-ions, or the bond geometries. It appears that the inter-layer ligands may have an unpredictable effect on the intra-layer magnetism -- given that we cannot account for the differences between any of the compounds based on the homologous geometric elements.

This work suggests that controlling the magnetic interactions between neighboring Cu atoms along Cu-pyz-Cu bonds will be extremely difficult. Small structural changes, e.g., the nature and position of the inter-layer ligands, lead to large changes in the relative strengths of different interactions within the same material. This would be extremely hard to predict prior to the synthesis of a new material. Unfortunately, our conclusions do not extend well to predicting or controlling the strength of magnetic interactions via Cu-pyz-Cu bonds in general. The effort to achieve designability through customizable geometry, in this case, has had an adverse affect because of the sensitivity of the interactions due to non-homologous features.

\textbf{Acknowledgments} The authors thank Ross McKenzie and Klaus Koepernik for helpful conversations. 
This work was supported by the Australian Research Council through Grant No. DP180101483.

\bibliography{goddard}

\begin{thebibliography}{27}%
\makeatletter
\providecommand \@ifxundefined [1]{%
 \@ifx{#1\undefined}
}%
\providecommand \@ifnum [1]{%
 \ifnum #1\expandafter \@firstoftwo
 \else \expandafter \@secondoftwo
 \fi
}%
\providecommand \@ifx [1]{%
 \ifx #1\expandafter \@firstoftwo
 \else \expandafter \@secondoftwo
 \fi
}%
\providecommand \natexlab [1]{#1}%
\providecommand \enquote  [1]{``#1''}%
\providecommand \bibnamefont  [1]{#1}%
\providecommand \bibfnamefont [1]{#1}%
\providecommand \citenamefont [1]{#1}%
\providecommand \href@noop [0]{\@secondoftwo}%
\providecommand \href [0]{\begingroup \@sanitize@url \@href}%
\providecommand \@href[1]{\@@startlink{#1}\@@href}%
\providecommand \@@href[1]{\endgroup#1\@@endlink}%
\providecommand \@sanitize@url [0]{\catcode `\\12\catcode `\$12\catcode
  `\&12\catcode `\#12\catcode `\^12\catcode `\_12\catcode `\%12\relax}%
\providecommand \@@startlink[1]{}%
\providecommand \@@endlink[0]{}%
\providecommand \url  [0]{\begingroup\@sanitize@url \@url }%
\providecommand \@url [1]{\endgroup\@href {#1}{\urlprefix }}%
\providecommand \urlprefix  [0]{URL }%
\providecommand \Eprint [0]{\href }%
\providecommand \doibase [0]{http://dx.doi.org/}%
\providecommand \selectlanguage [0]{\@gobble}%
\providecommand \bibinfo  [0]{\@secondoftwo}%
\providecommand \bibfield  [0]{\@secondoftwo}%
\providecommand \translation [1]{[#1]}%
\providecommand \BibitemOpen [0]{}%
\providecommand \bibitemStop [0]{}%
\providecommand \bibitemNoStop [0]{.\EOS\space}%
\providecommand \EOS [0]{\spacefactor3000\relax}%
\providecommand \BibitemShut  [1]{\csname bibitem#1\endcsname}%
\let\auto@bib@innerbib\@empty
\bibitem [{\citenamefont {Powell}(2020)}]{ContempPhys}%
  \BibitemOpen
  \bibfield  {author} {\bibinfo {author} {\bibfnamefont {B.~J.}\ \bibnamefont
  {Powell}},\ }\href {\doibase 10.1080/00107514.2020.1832350} {\bibfield
  {journal} {\bibinfo  {journal} {Contemp. Phys.}\ ,\ \bibinfo {pages} {1}}
  (\bibinfo {year} {2020})}\BibitemShut {NoStop}%
\bibitem [{\citenamefont {Kosterlitz}\ and\ \citenamefont
  {Thouless}(1973)}]{KT}%
  \BibitemOpen
  \bibfield  {author} {\bibinfo {author} {\bibfnamefont {J.~M.}\ \bibnamefont
  {Kosterlitz}}\ and\ \bibinfo {author} {\bibfnamefont {D.~J.}\ \bibnamefont
  {Thouless}},\ }\href {\doibase 10.1088/0022-3719/6/7/010} {\bibfield
  {journal} {\bibinfo  {journal} {J. Phys. C}\ }\textbf {\bibinfo {volume}
  {6}},\ \bibinfo {pages} {1181} (\bibinfo {year} {1973})}\BibitemShut
  {NoStop}%
\bibitem [{\citenamefont {Powell}\ and\ \citenamefont {McKenzie}(2011)}]{RPP}%
  \BibitemOpen
  \bibfield  {author} {\bibinfo {author} {\bibfnamefont {B.~J.}\ \bibnamefont
  {Powell}}\ and\ \bibinfo {author} {\bibfnamefont {R.~H.}\ \bibnamefont
  {McKenzie}},\ }\href {\doibase 10.1088/0034-4885/74/5/056501} {\bibfield
  {journal} {\bibinfo  {journal} {Rep. Prog. Phys.}\ }\textbf {\bibinfo
  {volume} {74}},\ \bibinfo {pages} {056501} (\bibinfo {year}
  {2011})}\BibitemShut {NoStop}%
\bibitem [{\citenamefont {Hassan}\ \emph {et~al.}(2018)\citenamefont {Hassan},
  \citenamefont {Cunningham}, \citenamefont {Mourigal}, \citenamefont
  {Zhilyaeva}, \citenamefont {Torunova}, \citenamefont {Lyubovskaya},
  \citenamefont {Schlueter},\ and\ \citenamefont {Drichko}}]{Drichko}%
  \BibitemOpen
  \bibfield  {author} {\bibinfo {author} {\bibfnamefont {N.}~\bibnamefont
  {Hassan}}, \bibinfo {author} {\bibfnamefont {S.}~\bibnamefont {Cunningham}},
  \bibinfo {author} {\bibfnamefont {M.}~\bibnamefont {Mourigal}}, \bibinfo
  {author} {\bibfnamefont {E.~I.}\ \bibnamefont {Zhilyaeva}}, \bibinfo {author}
  {\bibfnamefont {S.~A.}\ \bibnamefont {Torunova}}, \bibinfo {author}
  {\bibfnamefont {R.~N.}\ \bibnamefont {Lyubovskaya}}, \bibinfo {author}
  {\bibfnamefont {J.~A.}\ \bibnamefont {Schlueter}}, \ and\ \bibinfo {author}
  {\bibfnamefont {N.}~\bibnamefont {Drichko}},\ }\href {\doibase
  10.1126/science.aan6286} {\bibfield  {journal} {\bibinfo  {journal}
  {Science}\ }\textbf {\bibinfo {volume} {360}},\ \bibinfo {pages} {1101}
  (\bibinfo {year} {2018})}\BibitemShut {NoStop}%
\bibitem [{\citenamefont {Jacko}\ \emph {et~al.}(2020)\citenamefont {Jacko},
  \citenamefont {Kenny},\ and\ \citenamefont {Powell}}]{JackoPRM}%
  \BibitemOpen
  \bibfield  {author} {\bibinfo {author} {\bibfnamefont {A.~C.}\ \bibnamefont
  {Jacko}}, \bibinfo {author} {\bibfnamefont {E.~P.}\ \bibnamefont {Kenny}}, \
  and\ \bibinfo {author} {\bibfnamefont {B.~J.}\ \bibnamefont {Powell}},\
  }\href {\doibase 10.1103/PhysRevB.101.125110} {\bibfield  {journal} {\bibinfo
   {journal} {Phys. Rev. B}\ }\textbf {\bibinfo {volume} {101}},\ \bibinfo
  {pages} {125110} (\bibinfo {year} {2020})}\BibitemShut {NoStop}%
\bibitem [{\citenamefont {Kenny}\ \emph {et~al.}(2019)\citenamefont {Kenny},
  \citenamefont {Jacko},\ and\ \citenamefont {Powell}}]{KennyACIE}%
  \BibitemOpen
  \bibfield  {author} {\bibinfo {author} {\bibfnamefont {E.~P.}\ \bibnamefont
  {Kenny}}, \bibinfo {author} {\bibfnamefont {A.~C.}\ \bibnamefont {Jacko}}, \
  and\ \bibinfo {author} {\bibfnamefont {B.~J.}\ \bibnamefont {Powell}},\
  }\href {\doibase https://doi.org/10.1002/anie.201907889} {\bibfield
  {journal} {\bibinfo  {journal} {Angew. Chem. Int. Ed.}\ }\textbf {\bibinfo
  {volume} {58}},\ \bibinfo {pages} {15082} (\bibinfo {year}
  {2019})}\BibitemShut {NoStop}%
\bibitem [{\citenamefont {Batten}\ \emph {et~al.}(2009)\citenamefont {Batten},
  \citenamefont {Neville},\ and\ \citenamefont {Turner}}]{NevilleBook}%
  \BibitemOpen
  \bibfield  {author} {\bibinfo {author} {\bibfnamefont {S.}~\bibnamefont
  {Batten}}, \bibinfo {author} {\bibfnamefont {S.}~\bibnamefont {Neville}}, \
  and\ \bibinfo {author} {\bibfnamefont {D.}~\bibnamefont {Turner}},\
  }\href@noop {} {\emph {\bibinfo {title} {Coordination Polymers: Design,
  Analysis and Application}}}\ (\bibinfo  {publisher} {Royal Society of
  Chemistry},\ \bibinfo {year} {2009})\BibitemShut {NoStop}%
\bibitem [{\citenamefont {Robson}(2000)}]{Robson}%
  \BibitemOpen
  \bibfield  {author} {\bibinfo {author} {\bibfnamefont {R.}~\bibnamefont
  {Robson}},\ }\href {\doibase 10.1039/B003591M} {\bibfield  {journal}
  {\bibinfo  {journal} {J. Chem. Soc.{,} Dalton Trans.}\ ,\ \bibinfo {pages}
  {3735}} (\bibinfo {year} {2000})}\BibitemShut {NoStop}%
\bibitem [{\citenamefont {Goddard}\ \emph {et~al.}(2012)\citenamefont
  {Goddard}, \citenamefont {Manson}, \citenamefont {Singleton}, \citenamefont
  {Franke}, \citenamefont {Lancaster}, \citenamefont {Steele}, \citenamefont
  {Blundell}, \citenamefont {Baines}, \citenamefont {Pratt}, \citenamefont
  {McDonald}, \citenamefont {Ayala-Valenzuela}, \citenamefont {Corbey},
  \citenamefont {Southerland}, \citenamefont {Sengupta},\ and\ \citenamefont
  {Schlueter}}]{Goddard2012}%
  \BibitemOpen
  \bibfield  {author} {\bibinfo {author} {\bibfnamefont {P.~A.}\ \bibnamefont
  {Goddard}}, \bibinfo {author} {\bibfnamefont {J.~L.}\ \bibnamefont {Manson}},
  \bibinfo {author} {\bibfnamefont {J.}~\bibnamefont {Singleton}}, \bibinfo
  {author} {\bibfnamefont {I.}~\bibnamefont {Franke}}, \bibinfo {author}
  {\bibfnamefont {T.}~\bibnamefont {Lancaster}}, \bibinfo {author}
  {\bibfnamefont {A.~J.}\ \bibnamefont {Steele}}, \bibinfo {author}
  {\bibfnamefont {S.~J.}\ \bibnamefont {Blundell}}, \bibinfo {author}
  {\bibfnamefont {C.}~\bibnamefont {Baines}}, \bibinfo {author} {\bibfnamefont
  {F.~L.}\ \bibnamefont {Pratt}}, \bibinfo {author} {\bibfnamefont {R.~D.}\
  \bibnamefont {McDonald}}, \bibinfo {author} {\bibfnamefont {O.~E.}\
  \bibnamefont {Ayala-Valenzuela}}, \bibinfo {author} {\bibfnamefont {J.~F.}\
  \bibnamefont {Corbey}}, \bibinfo {author} {\bibfnamefont {H.~I.}\
  \bibnamefont {Southerland}}, \bibinfo {author} {\bibfnamefont
  {P.}~\bibnamefont {Sengupta}}, \ and\ \bibinfo {author} {\bibfnamefont
  {J.~A.}\ \bibnamefont {Schlueter}},\ }\href {\doibase
  10.1103/PhysRevLett.108.077208} {\bibfield  {journal} {\bibinfo  {journal}
  {Phys. Rev. Lett.}\ }\textbf {\bibinfo {volume} {108}},\ \bibinfo {pages}
  {077208} (\bibinfo {year} {2012})}\BibitemShut {NoStop}%
\bibitem [{\citenamefont {Landee}\ and\ \citenamefont
  {Turnbull}(2013)}]{Landee2013}%
  \BibitemOpen
  \bibfield  {author} {\bibinfo {author} {\bibfnamefont {C.~P.}\ \bibnamefont
  {Landee}}\ and\ \bibinfo {author} {\bibfnamefont {M.~M.}\ \bibnamefont
  {Turnbull}},\ }\href {\doibase 10.1002/ejic.201300133} {\bibfield  {journal}
  {\bibinfo  {journal} {European Journal of Inorganic Chemistry}\ }\textbf
  {\bibinfo {volume} {2013}},\ \bibinfo {pages} {2266} (\bibinfo {year}
  {2013})}\BibitemShut {NoStop}%
\bibitem [{\citenamefont {Vela}\ \emph {et~al.}(2013)\citenamefont {Vela},
  \citenamefont {Jornet-Somoza}, \citenamefont {Turnbull}, \citenamefont
  {Feyerherm}, \citenamefont {Novoa},\ and\ \citenamefont {Deumal}}]{Vela2013}%
  \BibitemOpen
  \bibfield  {author} {\bibinfo {author} {\bibfnamefont {S.}~\bibnamefont
  {Vela}}, \bibinfo {author} {\bibfnamefont {J.}~\bibnamefont {Jornet-Somoza}},
  \bibinfo {author} {\bibfnamefont {M.~M.}\ \bibnamefont {Turnbull}}, \bibinfo
  {author} {\bibfnamefont {R.}~\bibnamefont {Feyerherm}}, \bibinfo {author}
  {\bibfnamefont {J.~J.}\ \bibnamefont {Novoa}}, \ and\ \bibinfo {author}
  {\bibfnamefont {M.}~\bibnamefont {Deumal}},\ }\href {\doibase
  10.1021/ic400712s} {\bibfield  {journal} {\bibinfo  {journal} {Inorganic
  Chemistry}\ }\textbf {\bibinfo {volume} {52}},\ \bibinfo {pages} {12923}
  (\bibinfo {year} {2013})}\BibitemShut {NoStop}%
\bibitem [{\citenamefont {Goddard}\ \emph {et~al.}(2016)\citenamefont
  {Goddard}, \citenamefont {Singleton}, \citenamefont {Franke}, \citenamefont
  {M\"oller}, \citenamefont {Lancaster}, \citenamefont {Steele}, \citenamefont
  {Topping}, \citenamefont {Blundell}, \citenamefont {Pratt}, \citenamefont
  {Baines}, \citenamefont {Bendix}, \citenamefont {McDonald}, \citenamefont
  {Brambleby}, \citenamefont {Lees}, \citenamefont {Lapidus}, \citenamefont
  {Stephens}, \citenamefont {Twamley}, \citenamefont {Conner}, \citenamefont
  {Funk}, \citenamefont {Corbey}, \citenamefont {Tran}, \citenamefont
  {Schlueter},\ and\ \citenamefont {Manson}}]{Goddard2016}%
  \BibitemOpen
  \bibfield  {author} {\bibinfo {author} {\bibfnamefont {P.~A.}\ \bibnamefont
  {Goddard}}, \bibinfo {author} {\bibfnamefont {J.}~\bibnamefont {Singleton}},
  \bibinfo {author} {\bibfnamefont {I.}~\bibnamefont {Franke}}, \bibinfo
  {author} {\bibfnamefont {J.~S.}\ \bibnamefont {M\"oller}}, \bibinfo {author}
  {\bibfnamefont {T.}~\bibnamefont {Lancaster}}, \bibinfo {author}
  {\bibfnamefont {A.~J.}\ \bibnamefont {Steele}}, \bibinfo {author}
  {\bibfnamefont {C.~V.}\ \bibnamefont {Topping}}, \bibinfo {author}
  {\bibfnamefont {S.~J.}\ \bibnamefont {Blundell}}, \bibinfo {author}
  {\bibfnamefont {F.~L.}\ \bibnamefont {Pratt}}, \bibinfo {author}
  {\bibfnamefont {C.}~\bibnamefont {Baines}}, \bibinfo {author} {\bibfnamefont
  {J.}~\bibnamefont {Bendix}}, \bibinfo {author} {\bibfnamefont {R.~D.}\
  \bibnamefont {McDonald}}, \bibinfo {author} {\bibfnamefont {J.}~\bibnamefont
  {Brambleby}}, \bibinfo {author} {\bibfnamefont {M.~R.}\ \bibnamefont {Lees}},
  \bibinfo {author} {\bibfnamefont {S.~H.}\ \bibnamefont {Lapidus}}, \bibinfo
  {author} {\bibfnamefont {P.~W.}\ \bibnamefont {Stephens}}, \bibinfo {author}
  {\bibfnamefont {B.~W.}\ \bibnamefont {Twamley}}, \bibinfo {author}
  {\bibfnamefont {M.~M.}\ \bibnamefont {Conner}}, \bibinfo {author}
  {\bibfnamefont {K.}~\bibnamefont {Funk}}, \bibinfo {author} {\bibfnamefont
  {J.~F.}\ \bibnamefont {Corbey}}, \bibinfo {author} {\bibfnamefont {H.~E.}\
  \bibnamefont {Tran}}, \bibinfo {author} {\bibfnamefont {J.~A.}\ \bibnamefont
  {Schlueter}}, \ and\ \bibinfo {author} {\bibfnamefont {J.~L.}\ \bibnamefont
  {Manson}},\ }\href {\doibase 10.1103/PhysRevB.93.094430} {\bibfield
  {journal} {\bibinfo  {journal} {Phys. Rev. B}\ }\textbf {\bibinfo {volume}
  {93}},\ \bibinfo {pages} {094430} (\bibinfo {year} {2016})}\BibitemShut
  {NoStop}%
\bibitem [{\citenamefont {Tsyrulin}\ \emph {et~al.}(2010)\citenamefont
  {Tsyrulin}, \citenamefont {Xiao}, \citenamefont {Schneidewind}, \citenamefont
  {Link}, \citenamefont {R\o{}nnow}, \citenamefont {Gavilano}, \citenamefont
  {Landee}, \citenamefont {Turnbull},\ and\ \citenamefont
  {Kenzelmann}}]{Tsyrulin2010}%
  \BibitemOpen
  \bibfield  {author} {\bibinfo {author} {\bibfnamefont {N.}~\bibnamefont
  {Tsyrulin}}, \bibinfo {author} {\bibfnamefont {F.}~\bibnamefont {Xiao}},
  \bibinfo {author} {\bibfnamefont {A.}~\bibnamefont {Schneidewind}}, \bibinfo
  {author} {\bibfnamefont {P.}~\bibnamefont {Link}}, \bibinfo {author}
  {\bibfnamefont {H.~M.}\ \bibnamefont {R\o{}nnow}}, \bibinfo {author}
  {\bibfnamefont {J.}~\bibnamefont {Gavilano}}, \bibinfo {author}
  {\bibfnamefont {C.~P.}\ \bibnamefont {Landee}}, \bibinfo {author}
  {\bibfnamefont {M.~M.}\ \bibnamefont {Turnbull}}, \ and\ \bibinfo {author}
  {\bibfnamefont {M.}~\bibnamefont {Kenzelmann}},\ }\href {\doibase
  10.1103/PhysRevB.81.134409} {\bibfield  {journal} {\bibinfo  {journal} {Phys.
  Rev. B}\ }\textbf {\bibinfo {volume} {81}},\ \bibinfo {pages} {134409}
  (\bibinfo {year} {2010})}\BibitemShut {NoStop}%
\bibitem [{\citenamefont {Tsyrulin}\ \emph {et~al.}(2009)\citenamefont
  {Tsyrulin}, \citenamefont {Pardini}, \citenamefont {Singh}, \citenamefont
  {Xiao}, \citenamefont {Link}, \citenamefont {Schneidewind}, \citenamefont
  {Hiess}, \citenamefont {Landee}, \citenamefont {Turnbull},\ and\
  \citenamefont {Kenzelmann}}]{Tsyrulin2009}%
  \BibitemOpen
  \bibfield  {author} {\bibinfo {author} {\bibfnamefont {N.}~\bibnamefont
  {Tsyrulin}}, \bibinfo {author} {\bibfnamefont {T.}~\bibnamefont {Pardini}},
  \bibinfo {author} {\bibfnamefont {R.~R.~P.}\ \bibnamefont {Singh}}, \bibinfo
  {author} {\bibfnamefont {F.}~\bibnamefont {Xiao}}, \bibinfo {author}
  {\bibfnamefont {P.}~\bibnamefont {Link}}, \bibinfo {author} {\bibfnamefont
  {A.}~\bibnamefont {Schneidewind}}, \bibinfo {author} {\bibfnamefont
  {A.}~\bibnamefont {Hiess}}, \bibinfo {author} {\bibfnamefont {C.~P.}\
  \bibnamefont {Landee}}, \bibinfo {author} {\bibfnamefont {M.~M.}\
  \bibnamefont {Turnbull}}, \ and\ \bibinfo {author} {\bibfnamefont
  {M.}~\bibnamefont {Kenzelmann}},\ }\href {\doibase
  10.1103/PhysRevLett.102.197201} {\bibfield  {journal} {\bibinfo  {journal}
  {Phys. Rev. Lett.}\ }\textbf {\bibinfo {volume} {102}},\ \bibinfo {pages}
  {197201} (\bibinfo {year} {2009})}\BibitemShut {NoStop}%
\bibitem [{\citenamefont {Darriet}\ \emph {et~al.}(1979)\citenamefont
  {Darriet}, \citenamefont {Haddad}, \citenamefont {Duesler},\ and\
  \citenamefont {Hendrickson}}]{Darriet1979}%
  \BibitemOpen
  \bibfield  {author} {\bibinfo {author} {\bibfnamefont {J.}~\bibnamefont
  {Darriet}}, \bibinfo {author} {\bibfnamefont {M.~S.}\ \bibnamefont {Haddad}},
  \bibinfo {author} {\bibfnamefont {E.~N.}\ \bibnamefont {Duesler}}, \ and\
  \bibinfo {author} {\bibfnamefont {D.~N.}\ \bibnamefont {Hendrickson}},\
  }\href {\doibase 10.1021/ic50200a008} {\bibfield  {journal} {\bibinfo
  {journal} {Inorg. Chem.}\ }\textbf {\bibinfo {volume} {18}},\ \bibinfo
  {pages} {2679} (\bibinfo {year} {1979})}\BibitemShut {NoStop}%
\bibitem [{\citenamefont {Lancaster}\ \emph {et~al.}(2007)\citenamefont
  {Lancaster}, \citenamefont {Blundell}, \citenamefont {Brooks}, \citenamefont
  {Baker}, \citenamefont {Pratt}, \citenamefont {Manson}, \citenamefont
  {Conner}, \citenamefont {Xiao}, \citenamefont {Landee}, \citenamefont
  {Chaves}, \citenamefont {Soriano}, \citenamefont {Novak}, \citenamefont
  {Papageorgiou}, \citenamefont {Bianchi}, \citenamefont {Herrmannsd\"orfer},
  \citenamefont {Wosnitza},\ and\ \citenamefont {Schlueter}}]{Lancaster2007}%
  \BibitemOpen
  \bibfield  {author} {\bibinfo {author} {\bibfnamefont {T.}~\bibnamefont
  {Lancaster}}, \bibinfo {author} {\bibfnamefont {S.~J.}\ \bibnamefont
  {Blundell}}, \bibinfo {author} {\bibfnamefont {M.~L.}\ \bibnamefont
  {Brooks}}, \bibinfo {author} {\bibfnamefont {P.~J.}\ \bibnamefont {Baker}},
  \bibinfo {author} {\bibfnamefont {F.~L.}\ \bibnamefont {Pratt}}, \bibinfo
  {author} {\bibfnamefont {J.~L.}\ \bibnamefont {Manson}}, \bibinfo {author}
  {\bibfnamefont {M.~M.}\ \bibnamefont {Conner}}, \bibinfo {author}
  {\bibfnamefont {F.}~\bibnamefont {Xiao}}, \bibinfo {author} {\bibfnamefont
  {C.~P.}\ \bibnamefont {Landee}}, \bibinfo {author} {\bibfnamefont {F.~A.}\
  \bibnamefont {Chaves}}, \bibinfo {author} {\bibfnamefont {S.}~\bibnamefont
  {Soriano}}, \bibinfo {author} {\bibfnamefont {M.~A.}\ \bibnamefont {Novak}},
  \bibinfo {author} {\bibfnamefont {T.~P.}\ \bibnamefont {Papageorgiou}},
  \bibinfo {author} {\bibfnamefont {A.~D.}\ \bibnamefont {Bianchi}}, \bibinfo
  {author} {\bibfnamefont {T.}~\bibnamefont {Herrmannsd\"orfer}}, \bibinfo
  {author} {\bibfnamefont {J.}~\bibnamefont {Wosnitza}}, \ and\ \bibinfo
  {author} {\bibfnamefont {J.~A.}\ \bibnamefont {Schlueter}},\ }\href {\doibase
  10.1103/PhysRevB.75.094421} {\bibfield  {journal} {\bibinfo  {journal} {Phys.
  Rev. B}\ }\textbf {\bibinfo {volume} {75}},\ \bibinfo {pages} {094421}
  (\bibinfo {year} {2007})}\BibitemShut {NoStop}%
\bibitem [{\citenamefont {Barbero}\ \emph {et~al.}(2019)\citenamefont
  {Barbero}, \citenamefont {Medarde}, \citenamefont {Shang}, \citenamefont
  {Sheptyakov}, \citenamefont {Landee}, \citenamefont {Mesot}, \citenamefont
  {Ott},\ and\ \citenamefont {Shiroka}}]{Barbero2019}%
  \BibitemOpen
  \bibfield  {author} {\bibinfo {author} {\bibfnamefont {N.}~\bibnamefont
  {Barbero}}, \bibinfo {author} {\bibfnamefont {M.}~\bibnamefont {Medarde}},
  \bibinfo {author} {\bibfnamefont {T.}~\bibnamefont {Shang}}, \bibinfo
  {author} {\bibfnamefont {D.}~\bibnamefont {Sheptyakov}}, \bibinfo {author}
  {\bibfnamefont {C.~P.}\ \bibnamefont {Landee}}, \bibinfo {author}
  {\bibfnamefont {J.}~\bibnamefont {Mesot}}, \bibinfo {author} {\bibfnamefont
  {H.-R.}\ \bibnamefont {Ott}}, \ and\ \bibinfo {author} {\bibfnamefont
  {T.}~\bibnamefont {Shiroka}},\ }\href {\doibase
  10.1103/PhysRevMaterials.3.053602} {\bibfield  {journal} {\bibinfo  {journal}
  {Phys. Rev. Materials}\ }\textbf {\bibinfo {volume} {3}},\ \bibinfo {pages}
  {053602} (\bibinfo {year} {2019})}\BibitemShut {NoStop}%
\bibitem [{\citenamefont {Barbero}\ \emph {et~al.}(2016)\citenamefont
  {Barbero}, \citenamefont {Shiroka}, \citenamefont {Landee}, \citenamefont
  {Pikulski}, \citenamefont {Ott},\ and\ \citenamefont {Mesot}}]{Barbero2016}%
  \BibitemOpen
  \bibfield  {author} {\bibinfo {author} {\bibfnamefont {N.}~\bibnamefont
  {Barbero}}, \bibinfo {author} {\bibfnamefont {T.}~\bibnamefont {Shiroka}},
  \bibinfo {author} {\bibfnamefont {C.~P.}\ \bibnamefont {Landee}}, \bibinfo
  {author} {\bibfnamefont {M.}~\bibnamefont {Pikulski}}, \bibinfo {author}
  {\bibfnamefont {H.-R.}\ \bibnamefont {Ott}}, \ and\ \bibinfo {author}
  {\bibfnamefont {J.}~\bibnamefont {Mesot}},\ }\href {\doibase
  10.1103/PhysRevB.93.054425} {\bibfield  {journal} {\bibinfo  {journal} {Phys.
  Rev. B}\ }\textbf {\bibinfo {volume} {93}},\ \bibinfo {pages} {054425}
  (\bibinfo {year} {2016})}\BibitemShut {NoStop}%
\bibitem [{\citenamefont {Dos~Santos}\ \emph {et~al.}(2016)\citenamefont
  {Dos~Santos}, \citenamefont {Lanza}, \citenamefont {Barton}, \citenamefont
  {Brambleby}, \citenamefont {Blackmore}, \citenamefont {Goddard},
  \citenamefont {Xiao}, \citenamefont {Williams}, \citenamefont {Lancaster},
  \citenamefont {Pratt}, \citenamefont {Blundell}, \citenamefont {Singleton},
  \citenamefont {Manson},\ and\ \citenamefont {Macchi}}]{DosSantos2016}%
  \BibitemOpen
  \bibfield  {author} {\bibinfo {author} {\bibfnamefont {L.~H.~R.}\
  \bibnamefont {Dos~Santos}}, \bibinfo {author} {\bibfnamefont
  {A.}~\bibnamefont {Lanza}}, \bibinfo {author} {\bibfnamefont {A.~M.}\
  \bibnamefont {Barton}}, \bibinfo {author} {\bibfnamefont {J.}~\bibnamefont
  {Brambleby}}, \bibinfo {author} {\bibfnamefont {W.~J.~A.}\ \bibnamefont
  {Blackmore}}, \bibinfo {author} {\bibfnamefont {P.~A.}\ \bibnamefont
  {Goddard}}, \bibinfo {author} {\bibfnamefont {F.}~\bibnamefont {Xiao}},
  \bibinfo {author} {\bibfnamefont {R.~C.}\ \bibnamefont {Williams}}, \bibinfo
  {author} {\bibfnamefont {T.}~\bibnamefont {Lancaster}}, \bibinfo {author}
  {\bibfnamefont {F.~L.}\ \bibnamefont {Pratt}}, \bibinfo {author}
  {\bibfnamefont {S.~J.}\ \bibnamefont {Blundell}}, \bibinfo {author}
  {\bibfnamefont {J.}~\bibnamefont {Singleton}}, \bibinfo {author}
  {\bibfnamefont {J.~L.}\ \bibnamefont {Manson}}, \ and\ \bibinfo {author}
  {\bibfnamefont {P.}~\bibnamefont {Macchi}},\ }\href {\doibase
  10.1021/jacs.5b12817} {\bibfield  {journal} {\bibinfo  {journal} {J. Am.
  Chem. Soc.}\ }\textbf {\bibinfo {volume} {138}},\ \bibinfo {pages} {2280}
  (\bibinfo {year} {2016})}\BibitemShut {NoStop}%
\bibitem [{Note1()}]{Note1}%
  \BibitemOpen
  \bibinfo {note} {We maintain the labeling used by Goddard \protect \textit
  {et al.\protect \xspace }\ \cite {Goddard2016}.}\BibitemShut {Stop}%
\bibitem [{\citenamefont {Richardson}\ \emph {et~al.}(1977)\citenamefont
  {Richardson}, \citenamefont {Wasson},\ and\ \citenamefont
  {Hatfield}}]{Richardson1977}%
  \BibitemOpen
  \bibfield  {author} {\bibinfo {author} {\bibfnamefont {H.~W.}\ \bibnamefont
  {Richardson}}, \bibinfo {author} {\bibfnamefont {J.~R.}\ \bibnamefont
  {Wasson}}, \ and\ \bibinfo {author} {\bibfnamefont {W.~E.}\ \bibnamefont
  {Hatfield}},\ }\href {\doibase 10.1021/ic50168a053} {\bibfield  {journal}
  {\bibinfo  {journal} {Inorg. Chem.}\ }\textbf {\bibinfo {volume} {16}},\
  \bibinfo {pages} {484} (\bibinfo {year} {1977})}\BibitemShut {NoStop}%
\bibitem [{\citenamefont {Hatfield}\ and\ \citenamefont
  {Villa}(1971)}]{Hatfield1971}%
  \BibitemOpen
  \bibfield  {author} {\bibinfo {author} {\bibfnamefont {W.~E.}\ \bibnamefont
  {Hatfield}}\ and\ \bibinfo {author} {\bibfnamefont {J.~F.}\ \bibnamefont
  {Villa}},\ }\href {\doibase 10.1021/ja00745a062} {\bibfield  {journal}
  {\bibinfo  {journal} {J. Am. Chem. Soc.}\ }\textbf {\bibinfo {volume} {93}},\
  \bibinfo {pages} {4081} (\bibinfo {year} {1971})}\BibitemShut {NoStop}%
\bibitem [{\citenamefont {Mohri}\ \emph {et~al.}(1999)\citenamefont {Mohri},
  \citenamefont {Yoshizawa}, \citenamefont {Yambe}, \citenamefont {Ishida},\
  and\ \citenamefont {Nogami}}]{Mohri1999}%
  \BibitemOpen
  \bibfield  {author} {\bibinfo {author} {\bibfnamefont {F.}~\bibnamefont
  {Mohri}}, \bibinfo {author} {\bibfnamefont {K.}~\bibnamefont {Yoshizawa}},
  \bibinfo {author} {\bibfnamefont {T.}~\bibnamefont {Yambe}}, \bibinfo
  {author} {\bibfnamefont {T.}~\bibnamefont {Ishida}}, \ and\ \bibinfo {author}
  {\bibfnamefont {T.}~\bibnamefont {Nogami}},\ }\href {\doibase
  10.1023/A:1008300621000} {\bibfield  {journal} {\bibinfo  {journal} {Mol.
  Eng.}\ }\textbf {\bibinfo {volume} {8}},\ \bibinfo {pages} {357} (\bibinfo
  {year} {1999})}\BibitemShut {NoStop}%
\bibitem [{\citenamefont {Powell}(2011)}]{Reimers-chapter}%
  \BibitemOpen
  \bibfield  {author} {\bibinfo {author} {\bibfnamefont {B.~J.}\ \bibnamefont
  {Powell}},\ }\enquote {\bibinfo {title} {An introduction to effective
  low-energy hamiltonians in condensed matter physics and chemistry. {In JR
  Reimers (Ed.). C}omputational methods for large systems: Electronic structure
  approaches for biotechnology and nanotechnology},}\ \ (\bibinfo  {publisher}
  {Wiley},\ \bibinfo {address} {Hoboken},\ \bibinfo {year} {2011})\
  Chap.~\bibinfo {chapter} {10}, pp.\ \bibinfo {pages} {309--366}\BibitemShut
  {NoStop}%
\bibitem [{\citenamefont {Koepernik}\ and\ \citenamefont
  {Eschrig}(1999)}]{FPLO}%
  \BibitemOpen
  \bibfield  {author} {\bibinfo {author} {\bibfnamefont {K.}~\bibnamefont
  {Koepernik}}\ and\ \bibinfo {author} {\bibfnamefont {H.}~\bibnamefont
  {Eschrig}},\ }\href {\doibase 10.1103/PhysRevB.59.1743} {\bibfield  {journal}
  {\bibinfo  {journal} {Phys. Rev. B}\ }\textbf {\bibinfo {volume} {59}},\
  \bibinfo {pages} {1743} (\bibinfo {year} {1999})}\BibitemShut {NoStop}%
\bibitem [{\citenamefont {{de Graff}}\ and\ \citenamefont
  {Broer}(2016)}]{GraafBroer}%
  \BibitemOpen
  \bibfield  {author} {\bibinfo {author} {\bibfnamefont {C.}~\bibnamefont {{de
  Graff}}}\ and\ \bibinfo {author} {\bibfnamefont {R.}~\bibnamefont {Broer}},\
  }\href@noop {} {\emph {\bibinfo {title} {Magnetic Interactions in Molecules
  and Solids}}}\ (\bibinfo  {publisher} {Springer},\ \bibinfo {address}
  {Cham},\ \bibinfo {year} {2016})\BibitemShut {NoStop}%
\bibitem [{\citenamefont {Lu}\ \emph {et~al.}(2002)\citenamefont {Lu},
  \citenamefont {Xu}, \citenamefont {Wu}, \citenamefont {Wang},\ and\
  \citenamefont {Zhang}}]{Xin2002}%
  \BibitemOpen
  \bibfield  {author} {\bibinfo {author} {\bibfnamefont {X.}~\bibnamefont
  {Lu}}, \bibinfo {author} {\bibfnamefont {X.}~\bibnamefont {Xu}}, \bibinfo
  {author} {\bibfnamefont {J.}~\bibnamefont {Wu}}, \bibinfo {author}
  {\bibfnamefont {N.}~\bibnamefont {Wang}}, \ and\ \bibinfo {author}
  {\bibfnamefont {Q.}~\bibnamefont {Zhang}},\ }\href {\doibase
  10.1039/B105774J} {\bibfield  {journal} {\bibinfo  {journal} {New J. Chem.}\
  }\textbf {\bibinfo {volume} {26}},\ \bibinfo {pages} {160} (\bibinfo {year}
  {2002})}\BibitemShut {NoStop}%
\end{thebibliography}%

\end{document}


\title[]
	{Flexible magnetism in flexible crystals: Supporting Information}
	\author{E. P. Kenny}
	\affiliation{School of Mathematics and Physics, The University of Queensland, Brisbane, Queensland, Australia}
	\email{elisekenny@gmail.com}
	\author{A. C. Jacko}
	\affiliation{School of Mathematics and Physics, The University of Queensland, Brisbane, Queensland, Australia}
	\author{B. J. Powell}
	\affiliation{School of Mathematics and Physics, The University of Queensland, Brisbane, Queensland, Australia}
	
	\maketitle

\begin{figure*}
	\centering
	\includegraphics[width=\textwidth]{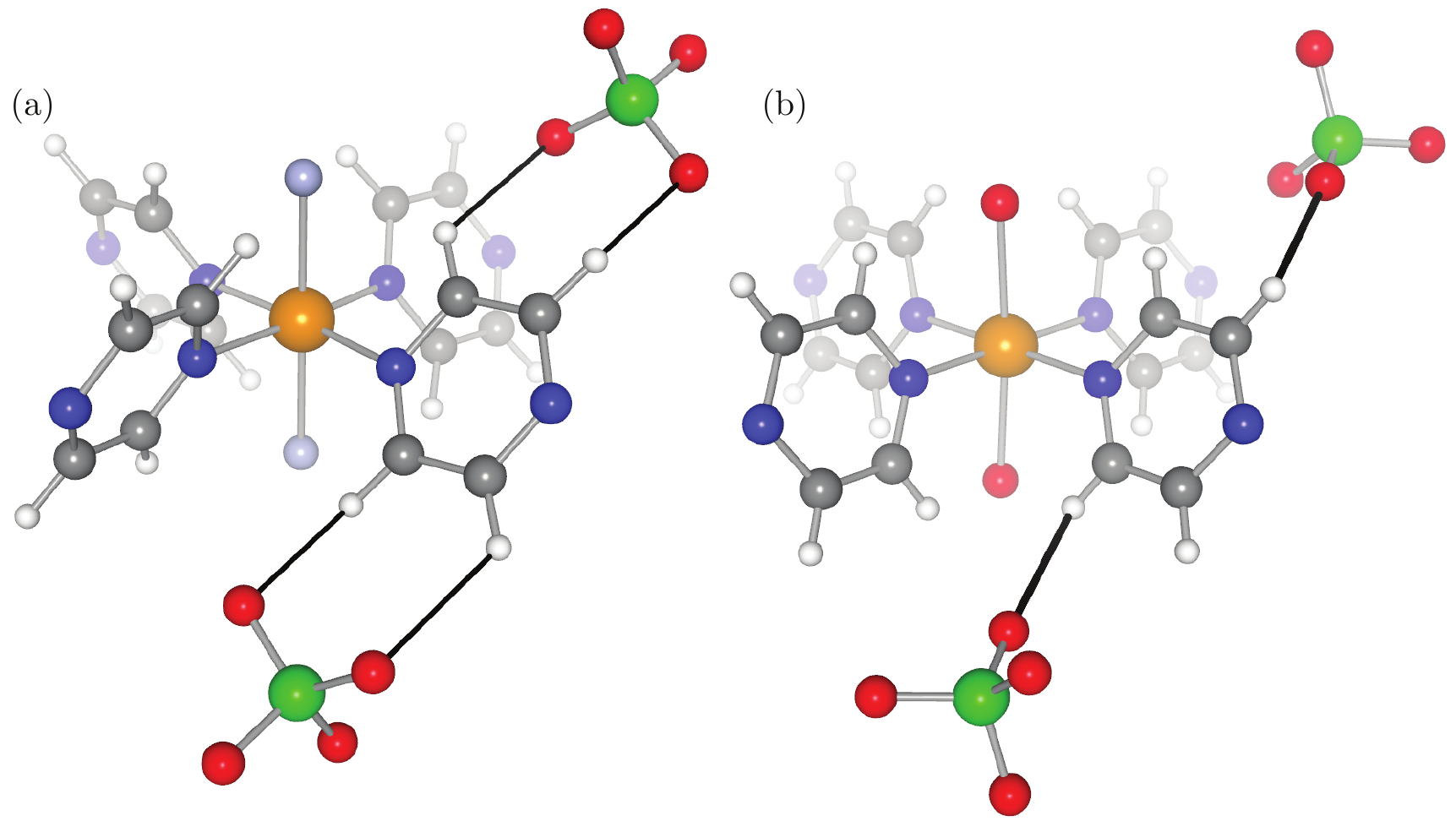}
	\caption{Examples of nearby H and O atoms used to measure the quantities in Figure 6 of the main text. Indicated with black lines, in (a) \textbf{1} and (b) \textbf{4}. }\label{fig:HO_indication}
\end{figure*}

\begin{figure*}
	\centering
	\includegraphics[width=\textwidth]{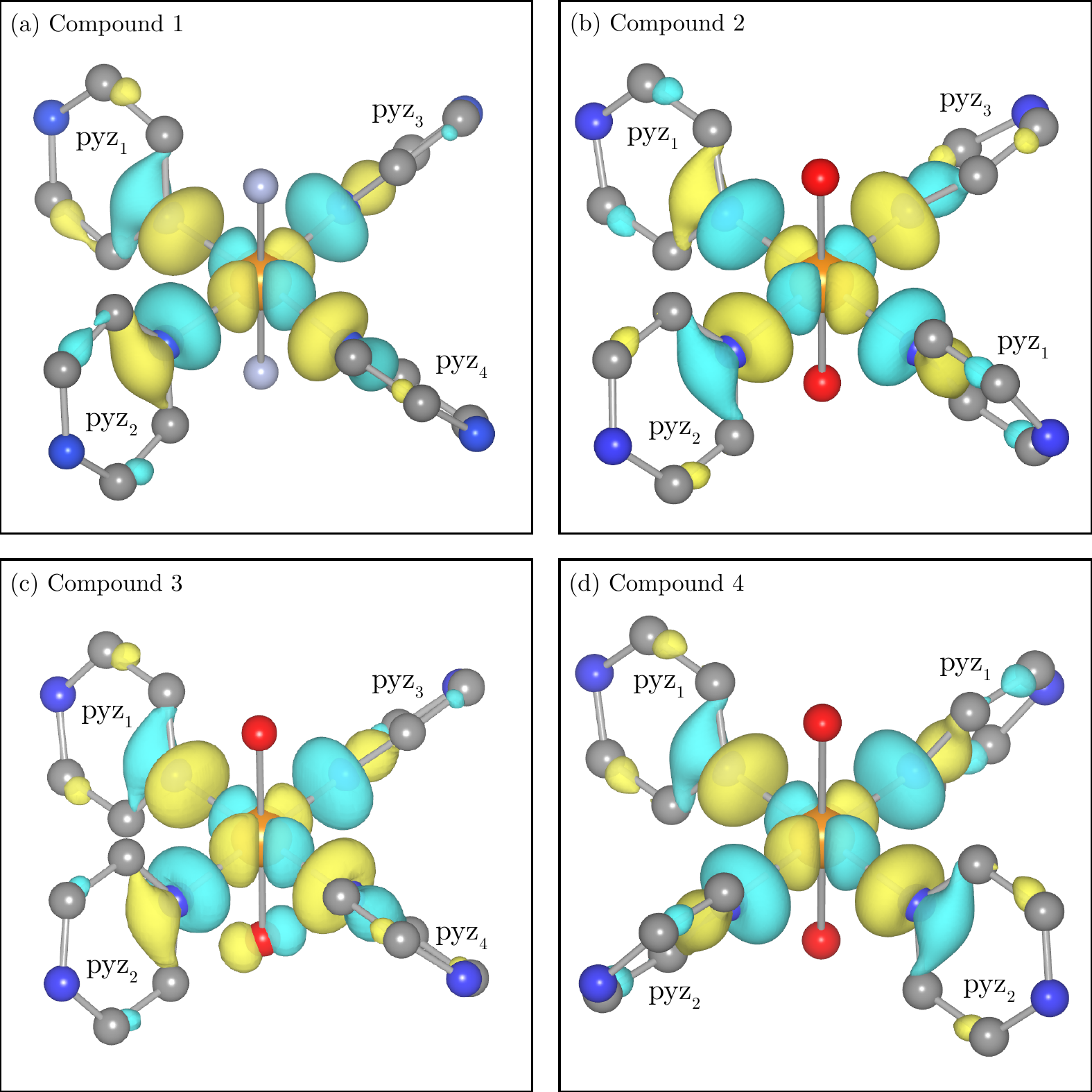}
	\caption{Cu-centered Wannier functions for the two-Wannier model (one on each Cu in the unit cell) for (a) \textbf{1}, (b) \textbf{2}, (c) \textbf{3}, and (d) \textbf{4}. All resemble hybrid orbitals of Cu d$_{x^2-y^2}$ and N sp$^2$. In \textbf{3}, there is some weight on one of the inter-layer ligands, 4-phpy-O.}\label{fig:all_wanniers_effective}
\end{figure*}

\begin{figure*}
	\centering
	\includegraphics[width=0.9\textwidth]{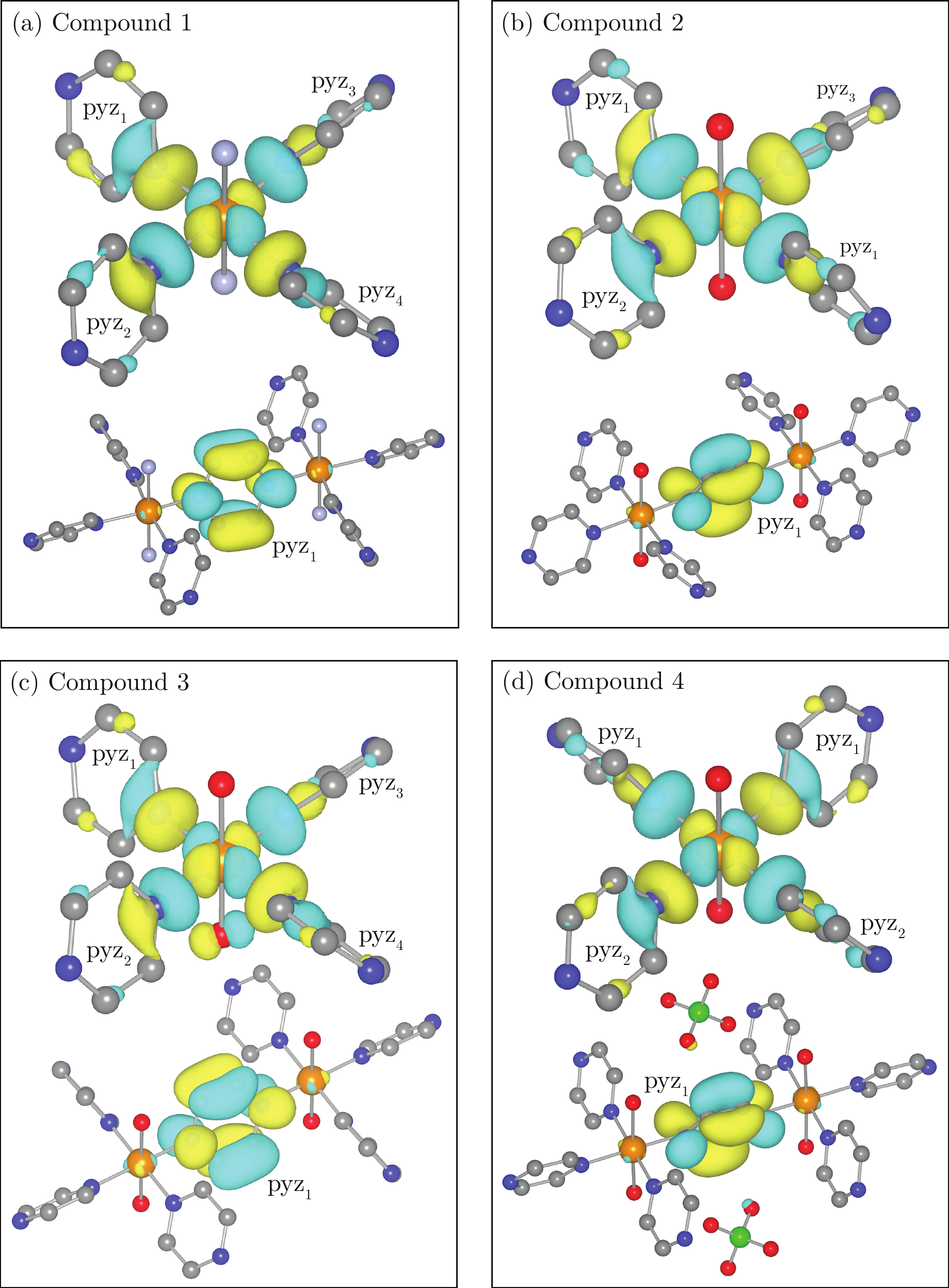}
	\caption{Cu-centered and pyz-centered Wannier functions for the six-Wannier model (one on each Cu and pyz in the unit cell) for (a) \textbf{1}, (b) \textbf{2}, (c) \textbf{3}, and (d) \textbf{4}. All Cu-centered functions are very similar to their counterparts in the two-Wannier model (in Figure \ref{fig:all_wanniers_effective}). The pyz-centered Wanniers all resemble the lowest, unoccupied molecular orbital of an isolated pyz. The other pyz-centered functions (not shown here) are all similar to the ones shown.}\label{fig:all_wanniers_pyz}
\end{figure*}
